\documentclass[pre,twocolumn,english,aps,prb,a4paper,floatfix]{revtex4}
\usepackage[T1]{fontenc}
\usepackage{tikz}
\usepackage{color}
\usepackage{amsmath}
\usepackage{amssymb}
\usepackage[latin1]{inputenc}
\usepackage{graphicx}
\usepackage{bm}
\usepackage{epsfig}
\usepackage{xcolor}
\usepackage{tcolorbox}
\usetikzlibrary{positioning}
\definecolor{mynicegreen}{RGB}{102,182,102}

\newcommand{\hypgeo}[2]{%
  \operatorname{%
    {\vphantom{\mathnormal{F}}}_{#1}%
    \kern-\scriptspace
    \mathnormal{F}_{#2}%
  }%
}

\begin{document}
\title{Density-functional theory for clustering of two-dimensional hard particle fluids}
\author{Yuri Mart\'{\i}nez-Rat\'on}
\email{yuri@math.uc3m.es}
\affiliation{
Grupo Interdisciplinar de Sistemas Complejos (GISC), Departamento
de Matem\'aticas, Escuela Polit\'ecnica Superior, Universidad Carlos III de Madrid,
Avenida de la Universidad 30, E-28911, Legan\'es, Madrid, Spain}

\author{Enrique Velasco}
\email{enrique.velasco@uam.es}
\affiliation{Departamento de F\'{\i}sica Te\'orica de la Materia Condensada,
Instituto de F\'{\i}sica de la Materia Condensada (IFIMAC) and Instituto de Ciencia de
Materiales Nicol\'as Cabrera,
Universidad Aut\'onoma de Madrid,
E-28049, Madrid, Spain}

\date{\today}

\begin{abstract}
 Fluids made of two-dimensional hard particles with polygonal shapes may stabilize symmetries which do not result directly from the particle shape. This is due to the formation of clusters in the fluid. Entropy alone can drive these effects, which represent a challenge for standard theories. In this article we present a general density-functional theory for clustering effects in fluids of hard particles in two dimensions. The theory combines a free-energy functional of the angular distribution
function with an association energy term which qualitatively reproduces the clustering tendencies of the particles found in Monte Carlo simulations. Application is made to a fluid of hard right-angled triangles.

\end{abstract}


\maketitle

\section{Introduction}

Hard particle models continue to attract great interest in the theory of phase transitions and colloidal physics \cite{review}. The hard-sphere model in two and three
dimensions demonstrated for the first time the existence of a crystal phase due solely to purely repulsive interactions \cite{HS_crystal1,HS_crystal2}, with important implications in colloidal science \cite{Poon}.
Anisotropic models have also proved useful to explain mesogenic formation involving both translational and orientational order \cite{Onsager,Parsons,Lee,Frenkel1,Frenkel2,Frenkel3,Frenkel4,Jack,Odrio1,Odrio2}.
The appearance of order in these systems is obviously due to entropic interactions driving the system to ordered or partially ordered configurations where the free volume available to the particles is maximized. The role of entropy in the formation of colloidal crystals has recently been emphasised and explained in terms of the concept of entropic bonding \cite{Glotzer0}.

In systems of identical particles, the extra volume available to each particle is usually shared equally among all the particles. However, in some systems this extra volume is not so trivially split \cite{GAN}. The configurational space is reorganized into internal coordinates, associated to the formation of clusters, which internally loose entropy, while the gain in entropy of collective coordinates associated to the cluster centers of mass more than compensates the former entropy loss. The net effect is the formation of differently shaped \textit{superparticles} which can favour the stability of high-order symmetries, not evident from the particle shape alone, in the fluid.

This effect may be especially important in two-dimensional (2D) systems made of polygonally-shaped particles. Some particle shapes seem to be more prone to exhibit these entropic clustering effects, which have
a huge impact on the stability of highly-symmetric fluid phases. The compartmentalization of entropy between low- and high-entropy subsystems has been invoked to design host-guest chemical systems
with important consequences for nanoparticle storage \cite{Glotzer1,Glotzer2} in colloidal crystals. A general theory to describe storage and encapsulation in equilibrium mixtures of tailored hard particles has been proposed recently \cite{Wittmann}. However, there is evidence that the entropy-driven self-assembly of identical hard particles into clusters is also at work in 2D one-component liquid-crystal, as opposed to crystal phases or mixtures \cite{MIG1,Proc}.

Many-body effects leading to self-assembly in 2D oriented fluids represent a challenge for standard theories for liquid crystals \cite{MAR2}. These theories, the most successfully of which is Scaled-Particle Theory (SPT), are based on the Onsager theory \cite{Onsager},
whose central ingredient is the volume (area) excluded to a particle by another particle. This is totally equivalent to considering a virial
expansion truncated at second order, i.e. at two-body correlations. However, the formation of clusters of particles necessarily involve
higher-order correlations. Some proposals of extended theories that incorporate these correlations systematically, in the context of 2D oriented fluids, have been made \cite{VEL1}.
The effect of third- and fourth-order correlations on the stability of the tetratic (T) phase in fluids of hard rectangles and
hard right-angled triangles have been analysed, with relatively disappointing results in the latter case \cite{MAR2}. In particular, Monte Carlo (MC) simulations reveal a T phase with strong octatic (eightfold or 8-atic, O)  correlations for right-angled triangles \cite{GAN,MAR3}, but
the extended theories only predict a uniaxial nematic (N) in the whole density range. 
Even a theory containing the exact fourth
virial coefficient, which would seem at first glance as an essential ingredient to capture the formation of square dimers and tetramers
(which are behind the stability of O fluctuations in the nematic director), fails completely in this respect \cite{VEL1}.
It must be said that the difference between the isotropic (I)-N and 
I-O bifurcation packing fractions tend to converge as more virial coefficients are included. Also, the corresponding theories involve calculations with a huge numerical burden.

Clearly a density-functional theory constructed from few-particle correlations does not have the correct structure to describe clustering and the relative equilibrium distributions of even the most basic particle arrangements.
A first step to explore how the new symmetries can be generated considers these clusters as fixed entities or superparticles. This view leads to an extreme model, where clusters may orient in such a way as to give rise to high-order symmetries in the final monomer orientational distribution function, e.g. the orientational eightfold symmetry of the O phase in the case of right-angled triangles \cite{MAR3}. However, the
fractions of the clusters have to be imposed by hand and are not obtained self-consistently.

\begin{figure}[h]
\begin{center}
        \includegraphics[width=0.90\linewidth,angle=0]{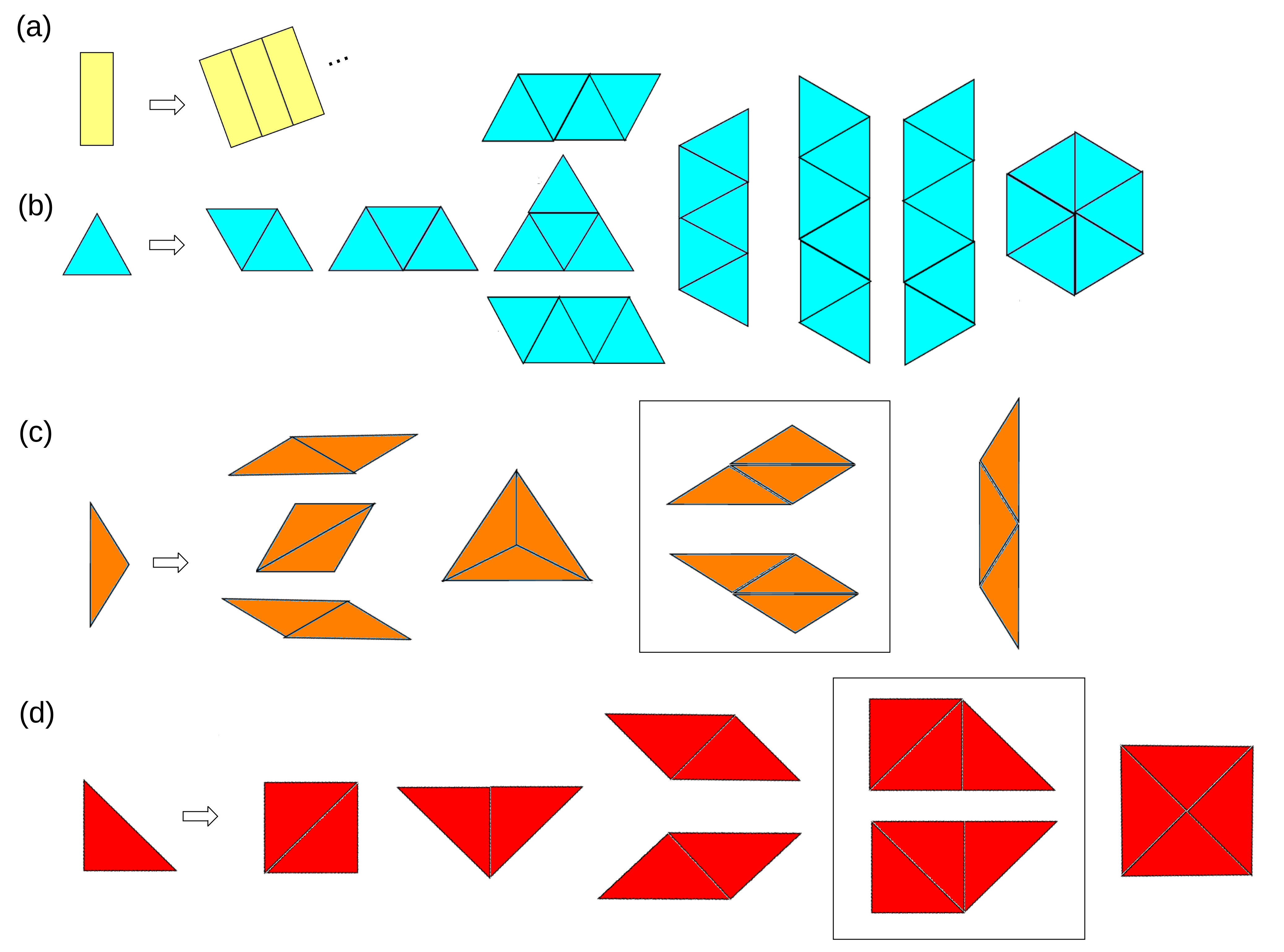}
        \caption{\label{fig1} Candidate hard polygonal particles to be described by the present theory for entropic
        clustering. On the left a single particle is depicted for each model. (a) Rectangles
        form rectangular clusters by joining particles side by side. For a given aspect ratio the most abundant
        cluster has a shape close to a square, which favours the T phase. (b) Equilateral triangles may form a
        variety of shapes, the most abundant ones being dimers and hexagons made of six particles. (c) Isosceles triangles with an opening angle of $120^{\circ}$ may form stable rhomboidal clusters made of two particles and equilateral triangular and trapezoidal clusters made of three particles.
        (d) Right-angles triangles form large variety of dimers, including two chiral rhombuses,
        and a tetramer with square symmetry. 
	Clusters enclosed in boxes are not intended to be included in the theory in a first implementation.}
\end{center}
\end{figure}

A number of particle models are clear candidates for exhibiting strong clustering effects in their liquid-crystalline and crystal phases. Both regular and nonregular
polygonal shapes are included in the list. As a general criterion, particles that can be associated into clusters
with shapes that tessellate the area can be considered to be potential candidates. The clustering tendencies of some of these
particles have already been explored previously. Fig. \ref{fig1} 
shows some candidate hard polygonal-particle models for which one would expect to find strong clustering: rectangles, equilateral triangles, isosceles triangles with an opening angle of $120^{\circ}$, and right-angled triangles. Several single particles can be combined into dimers, trimers, tetramers and larger clusters of different shapes. The correct criteria to a priori identify the important cluster shapes are not evident; however, a reasonable criterion should include an element of roundness and convexity that permits some rotational and translational freedom of the clusters. 

Some of these superparticles may be particularly stable in the equilibrium fluid and give a dominant contribution to the structure and free energy of the fluid, to the extent that the global symmetry of the bulk phase may be determined by the abundance of these clusters with particular symmetries. The global symmetry may be very different from the one that would directly follow from the shape of the particle. Particular clusters will be dominant at high densities and may even characterise uniquely the crystal phase. For example, rectangles with an aspect ratio of two will form square dimers, which crystallise into a high-entropy crystal, presumably formed by dimers oriented randomly. Likewise, equilateral triangles will tend to form hexagonal crystals made of hexamers with hexagonal shape \cite{GAN}. Finally, right-angled triangles form particularly stable tetramers with square shape, which crystallise into a square lattice \cite{GAN}.

Some of these systems have already been explored, either by theory, simulation or experiment. Hard rectangles of aspect ratio $\kappa$ have been studied using the excluded-volume-based SPT \cite{Schlacken,Effect} and higher-order virial
theories \cite{Third}, as well as by simulation \cite{Woj,Donev} and experiments on colloidal particles \cite{Chaikin} and on vibrated granular monolayers \cite{MIG1}. All of these studies have shown
a tendency for the formation of clusters of approximately square shape containing a number of particles close to $\kappa$.
These clusters orient at right angles and enhance the stability of the fourfold T (or 4-atic) phase over the I or disordered phase \cite{MAR1}.
Rectangular clusters of larger size, involving a larger number of parallel rectangles, are also
easily observed in simulations and experiments. The size distribution of these clusters decays exponentially with their size,
and extended SPT theories that consider these clusters to be different species have shown the square cluster to have the dominant contribution
in the free energy \cite{MAR3}. These clusters are the seeds for the formation of large layered regions with smectic symmetry \cite{Narayan,MIG1}.

Hard equilateral and right-angled triangles have been investigated by MC simulation by Gantapara et al. \cite{GAN}.
These particles not only exhibit strong clustering in the crystal phase, but also chiral features: particles associate into clusters that may rotate in one of two
possible senses to give rise to chiral configurations. For example, in the crystal phase, four right-angles triangles form tightly-bound square clusters that
behave as single superparticles \cite{explicacion} and collectively rotate in one direction to reduce their entropy, which compensates the entropy increase of the cluster. Since two rotations are possible, this is a clear example of chiral phases that coexist separated by interfaces \cite{GAN} in an otherwise achiral global phase.

\begin{figure}[h]
\begin{center}
        \includegraphics[width=0.90\linewidth,angle=0]{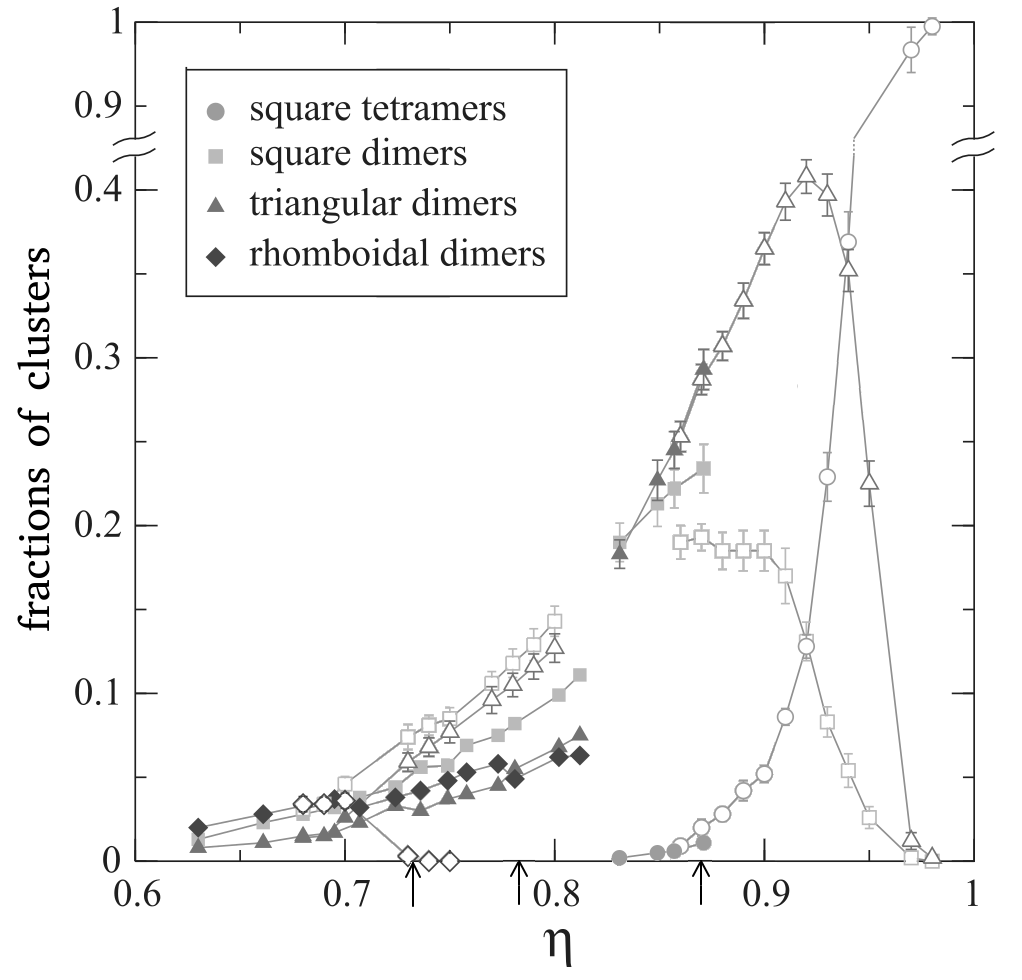}
        \caption{\label{nuevafig2} Fractions of clusters as a function of packing fraction in the system of right-angled triangles as obtained by MC simulation (adapted from \cite{MAR3}). Clusters are represented by different symbols (see key). Open symbols are expansion runs, while filled symbols correspond to compression runs. Arrows in the $\eta$ axis indicate the coexistence densities of I ($0.733$) and T ($0.782$), and the transition density to the crystal ($0.87$), according to Ref. \cite{GAN}.
        }
\end{center}
\end{figure}

The clustering properties of the fluid phases of right-angles triangles have also been theoretically studied in detail using extended SPT and
simulations \cite{MAR2,VEL1}. A 4-atic phase with strong 8-atic correlations predicted by MC simulations seems to be the most likely candidate for oriented fluid in this system.
The strong eightfold correlations can only be explained as a result of the presence of various types of square dimers, square tetramers, and rhomboidal and triangular dimers in the fluid, causing
predominant fourfold symmetries along with secondary eightfold symmetries 
(see Fig. \ref{nuevafig2} where fractions of the different types of clusters, obtained from
MC simulation, are represented
as a function of packing fraction \cite{MAR3}). These symmetries are not shared by the particle shape and result from the formation of
clusters with symmetries different from that of the individual particles.

\begin{figure}
\includegraphics[width=2.5in]{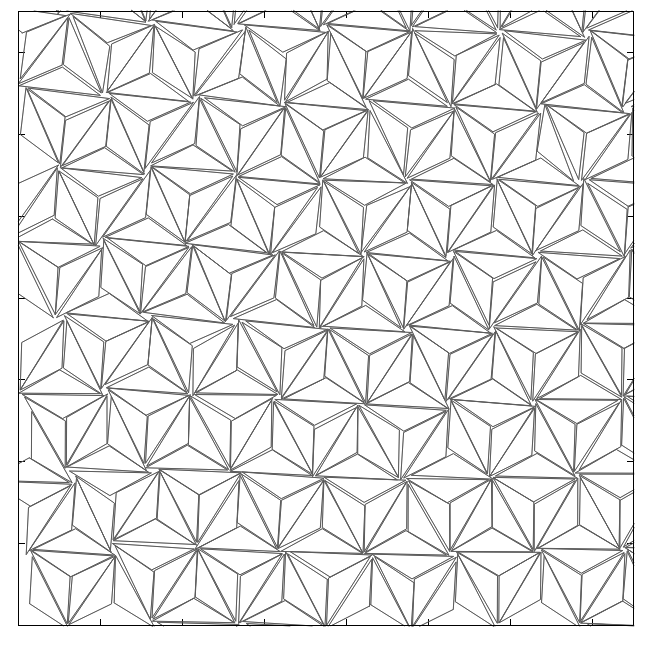}
	\caption{Particle configurations of the crystalline phase of isosceles triangles with an aperture angle of 120$^{\circ}$, obtained by Monte Carlo simulation. The equilateral triangular clusters formed by three 
triangles and with any neighboring clusters having opposite orientations ($180^{\circ}$ rotations) conform the crystal phase. Clusters rotate by a small angle, giving rise to locally chiral regions.}
\label{fig3}
\end{figure}

Many other possibilities exist. One involves isosceles triangles with an aperture angle of 120$^{\circ}$. Three such triangles form an equilateral triangle. At high density these trimers will stabilise a crystal with sixfold symmetry. Fig. \ref{fig3} shows a snapshot of a typical configuration of such a crystal, obtained from our own MC simulations (not previously published). Note that the partitioning of configurational space is revealed by local rotations of clusters, which leave an extra area in the interstitial spaces that increase the collective free area of clusters.  
At intermediate fluid densities a 6-atic, or triatic, phase can be stabilised. It is not unlikely that these clusters involving three particles
cannot be excited at lower densities, and that a N or 2-atic phase be the stable mesophase. Clearly subtle entropic effects, giving rise to unexpected phase diagrams, may be at work.

In this article we propose and explore a general theory, the Entropic Clustering Theory (ECT), that attempts to describe liquid-crystal phases with strong clustering tendencies in systems with arbitrary polygon-shaped particles. The theory combines a density-functional theoretical setting with standard approaches for associating fluids, and is inspired by previous applications to dipolar, polymer, hard-sphere, and patchy-fluid systems \cite{vanRoij,Segura,Cuesta,Tavares,delasHeras,Torres}. Orientational order is described through the usual angular distribution function, and a free-energy functional is written which includes two-body excluded-area effects. In addition, entropic tendencies of particles to cluster are accounted for by an association energy. Such an energy incorporates a penalty barrier against cluster formation, together with an effective coupling with the order parameters. As shown below, these ingredients are sufficient  to describe the clustering tendencies found in MC simulations. Sizes and shapes of clusters are defined a priori using common sense, while their statistical abundances in equilibrium result from a chemical-equilibrium condition. We combine these ideas with density-functional theory to describe both clustering and orientational ordering in 2D fluids. As an example, we apply the theory to a fluid made of right-angled triangles. 
To complete the analysis, a cell theory (CeT) is used to approximate the equation of state (EOS) of the crystal phase, made of tetramers located at the sites of a simple square lattice. As will be seen, the resulting theory accurately reproduces not only the EOS of the crystal phase, but also that of the T phase, when compared with simulation.

The remainder of the article is organised as follows. In Section \ref{Theory} we describe the theoretical framework, while results from the theory are shown in Section \ref{results}. Some conclusion and prospects for future applications of the theory to other particle shapes are drawn in Section \ref{conclusions}.

\section{Theory}
\label{Theory}

\subsection{Entropic clustering theory}

Despite the fact that the theory is general, we assume for definiteness that the fluid is made of hard right-angled triangular (HRT) particles, Fig. \ref{fig1}(d).
Our model consists of a mixture of six species which, apart from the triangular monomers, are 
formed by assembling monomeric units into five different clusters: square dimers, 
resulting by joining two HRT along their hypotenuses; triangular dimers, obtained by joining 
two HRT along their catets, so that the right angles of both triangles are in contact; two different 
rhomboidal dimers, one obtained from the other by specular reflection with respect to the common contact edge, giving rise 
to two different enantiomers (these rhomboids are formed by sticking two HRT through their catets but this 
time with the right angles being in contact with the 45$^{\circ}$-angles); and finally square tetramers, formed by joining four triangles together by their catets with their right-angle vertexes converging at the center of a perfect square. Note that there are two more clusters (also enantiomers) composed by three monomers (enclosed by a rectangular box in Fig. \ref{fig1}(d)) which, for the sake of simplicity, are not taken into account in our model. It is hoped that their effects do not significantly change the phase behavior of the system. 

Once again we stress the fact that our model is completely general, and valid for any multicomponent 
two-dimensional mixture of hard particles including monomers and a certain number of clusters obtained by self-assembling of monomers. The criterion to select the geometry of clusters is 
mainly dictated by their convexity. Certainly convex clusters will retain a much higher local entropy than nonconvex clusters, which may interlock with neighbours and reduce the entropy. However, this criterion also constitutes a convenient constraint which facilitates the calculation of excluded areas. Also, cluster fractions are variables in our model and it is convenient not to use too many species in order to make the calculations feasible. The central quantities that characterize the mixture are the density angular profiles $\rho_i(\phi)$ ($i=1,\dots,6$), with $\phi$ the angle between the particle axis and the polar axis of a fixed reference frame. 

Clusters interact through hard-core interactions and the corresponding interaction free-energy will be approximated by that obtained from scaled particle theory (SPT) extended to a multicomponent mixture \cite{Cotter1,Cotter2,nos}. 
The interaction or excess part of the free-energy density in reduced thermal units is therefore:
\begin{eqnarray}
	\Phi_{\rm int}[\{\rho_i\}]&&\equiv \frac{\beta {\cal F}_{\rm in}[\{\rho_i\}]}{A}\nonumber\\
	&&=-\rho \log(1-\eta)+\frac{\sum_{i,j} \langle \langle A_{ij}^{(\rm spt)}\rangle\rangle}{1-\eta},
\label{int}
\end{eqnarray}
where ${\cal F}_{\rm int}[\{\rho_i\}]$ is the interaction or excess part of the Helmholtz free-energy of the 
mixture and $A$ is the total area of the system. As usual $\beta$ is the inverse thermal energy. 
The quantity
\begin{eqnarray}
\eta\equiv\sum_i\rho_i a_i=\rho_0 a_0
\end{eqnarray}
is the total packing fraction of the mixture (packing fractions of monomers), and it is taken as a fixed number, equal to the total number density of monomers, $\rho_0$, multiplied by its area $a_0$. The quantity $\rho_i\equiv\int_0^{2\pi}d\phi \rho_i(\phi)$ 
is the number density of the cluster $i$, while $a_i$ is the area, equal to the product of the number of triangles forming the $i$-cluster, $n_i$, and the area of the monomers $a_0$: $a_i=n_ia_0$. The total number density of clusters is just the 
sum $\rho=\sum_i\rho_i$. Eqn. (\ref{int}) defines the double 
angular average of the spt-area $A_{ij}^{(\rm spt)}(\phi_1-\phi_2)$ with respect to 
the angular profiles $\rho_i(\phi_1)$ and $\rho_j(\phi_2)$:
\begin{eqnarray}
	&&\langle \langle A_{ij}^{(\rm spt)}\rangle \rangle=
    \int_0^{2\pi}d\phi_1\int_0^{2\pi}d\phi_2 \rho_i(\phi_1)\rho_j(\phi_2)\nonumber\\
	&&\times A_{ij}^{(\rm spt)}(\phi_1-\phi_2).
\label{double}
\end{eqnarray}
This area in turn is related with the excluded area as
\begin{eqnarray}
A_{ij}^{(\rm spt)}(\phi)=\frac{1}{2}\left(A_{ij}^{(\rm excl)}(\phi)-a_i-a_j\right).
\end{eqnarray}

We should note that if $N$ is the total number of components of the mixture (the total number of clusters plus one) we need to calculate a total of $N(N+1)/2$ different excluded
areas. In this case the number is 21. However there are some symmetries dictated by the interior angles of the polygonal clusters which allow to find some 
relations between the excluded areas. 

The ideal part of the multicomponent mixture is 
\begin{eqnarray}
\Phi_{\rm id}[\{\rho_i\}]\equiv \sum_i \int_0^{2\pi}d\phi \rho_i(\phi) \left[
\log(\rho_i(\phi))-1\right],
\end{eqnarray}
where the thermal areas have been dropped. Finally, an important part of the free-energy functional is 
a term involved in the balance of entropy when $n_i$ monomers assemble into a cluster of index $i$. We call this part the \textit{association energy} and make it to explicitly depend on the T order parameters $Q_4^{(i)}$:
\begin{eqnarray}
    \Phi_{\rm as}[\{\rho_i\}]\equiv\sum_i \rho_i \left[\epsilon_i^{(1)}+\epsilon_i^{(2)}\left(Q_4^{(i)}\right)^2\right],
    \label{association}
\end{eqnarray}
where $\epsilon_i^{(j)}$ are \textit{energylike} parameters (see discussion in Section \ref{parameters}).
This dependence is motivated in our present case by the favoured clusterization into tetramers when the T order parameter is high. For general particle symmetries the term enclosed by square brackets can be replaced by a general 
function 
$v_{\rm as}\left(\eta,\{Q_{2n}^{(i)}\}\right)$, which could also depend on the packing 
fraction and on the rest of order parameters. The order parameters of the $i$th species are in turn defined as the 
angular average of $\cos(2n\phi)$ with respect to the angular 
distribution function of the $i$th species,
\begin{eqnarray}
    h_i(\phi)\equiv \frac{\rho_i(\phi)}{\rho_i}\Rightarrow \int_0^{2\pi} d\phi 
    h_i(\phi)=1,
\end{eqnarray}
resulting in 
\begin{eqnarray}
Q_{2n}^{(i)}=\int_0^{2\pi}d\phi h_i(\phi)\cos(2n\phi),
\end{eqnarray}
where $n$ determines the symmetry of the phase: $n=2$ and $4$ for N and T symmetries, respectively. 

The equilibrium number-density angular profiles $\rho_i(\phi)$ are obtained from 
the ECT presented here by minimizing the constrained total free-energy density functional, 
\begin{eqnarray}
	&&\Phi(\{\rho_i\})=\Phi_{\rm id}(\{\rho_i\})+\Phi_{\rm int}(\{\rho_i\})+
\Phi_{\rm as}(\{\rho_i\})\nonumber\\
	&&-\lambda\left(\sum_i \rho_i n_i-\rho_0\right),
\end{eqnarray}
(with $\lambda$ the Lagrange multiplier) with respect to $\rho_i(\phi)$: 
$\displaystyle{\frac{\delta \Phi(\{\rho_i\})}{\delta \rho_i(\phi)}=0}$. This minimization 
results in 
\begin{eqnarray}
    \rho_i(\phi)=\rho_ih_i(\phi)=z^{n_i} e^{-c_1^{(i)}(\phi)},\quad z\equiv e^{\lambda},
    \label{la_rho}
\end{eqnarray}
where we have defined 
    \begin{eqnarray}
        &&c_1^{(i)}(\phi)=-\log(1-\eta)+\frac{{\cal S}_i(\phi)}
        {1-\eta}+\epsilon_i^{(1)}\nonumber
	    \\&&+\epsilon_i^{(2)}\left[2Q_4^{(i)}\cos(4\phi)-\left(Q_4^{(i)}\right)^2\right],\\
        &&{\cal S}_i(\phi)=2\sum_j \int_0^{2\pi} 
        d\phi' \rho_j(\phi') A_{ij}^{(\rm spt)}(\phi-\phi') \label{la_s}.
    \end{eqnarray}
Defining also the quantities 
\begin{eqnarray}
T_i\equiv \int_0^{2\pi} d\phi e^{-c_1^{(i)}(\phi)},
\end{eqnarray}
we obtain
\begin{eqnarray}
\rho_i=z^{n_i} T_i,
\label{set1}
\end{eqnarray}
and from the conservation number of monomers 
distributed among the clusters, we have 
\begin{eqnarray}
    \rho_0=\sum_i n_i T_i z^{n_i}.
    \label{poly}
\end{eqnarray}
In our case this is a fourth-order polynomial with respect 
to $z$ which is solved at each step of the numerical procedure 
to find the functions $\rho_i(\phi)$.

To perform the numerical minimization the angular distribution functions $h_i(\phi)$
are Fourier expanded,
\begin{eqnarray}
    h_i(\phi)=\frac{1}{2\pi}\left[1+\sum_{k\geq 1}h_k^{(i)}\cos(2k\phi)\right],
    \label{Fourier}
\end{eqnarray}
with $\{h_k^{(i)}\}$ the set of Fourier amplitudes. From Eqn. (\ref{double}) we obtain
\begin{eqnarray}
\langle \langle A_{ij}^{(\rm spt)}\rangle\rangle=\rho_i\rho_j\left[A_{ij,0}^{(\rm spt)}+\frac{1}{2}\sum_{k\geq 1}
A_{ij,k}^{(\rm spt)}h_k^{(i)}h_k^{(j)}\right],
\end{eqnarray}
and from Eqn. (\ref{la_s}),
\begin{eqnarray}
{\cal S}_i(\phi)=2\sum_j \rho_j\left[A_{ij,0}^{(\rm spt)}+\sum_{k\geq 1} A_{ij,k}^{(\rm spt)}h_k^{(j)}\cos(2k\phi)
\right],
\end{eqnarray}
where we have defined the Fourier components of the spt-areas as 
\begin{eqnarray}
A_{ij,k}^{(\rm spt)}\equiv \frac{1}{2\pi}\int_0^{2\pi} d\phi \cos(2k\phi) A_{ij}^{(\rm spt)}(\phi).
\end{eqnarray}
Inserting Eqn. (\ref{Fourier}) into Eqn. (\ref{la_rho}), multiplying by $\cos(2k\phi)$ and integrating over $\phi$, we obtain 
\begin{eqnarray}
	&& h_k^{(i)}=\frac{2\tilde{T}_i^{(k)}}{\tilde{T}_i^{(0)}},\ 
	\tilde{T}_i^{(k)}=\int_0^{2\pi}d\phi\cos(2k\phi) e^{- S_i(\phi)},\label{set2}\\ 
	&&S_i(\phi)=\frac{{\cal S}_i(\phi)}{1-\eta}+2\epsilon_i^{(2)}Q_4^{(i)}\cos(4\phi).
\end{eqnarray}
The unknowns are $z$, $\{\rho_i\}$, and $\{h_k^{(i)}\}$. Taking into account that 
$T_i$ and $\tilde{T}_i$ depend on the latter magnitudes, we use Picard iteration to solve the set 
of Eqns. (\ref{set1}) and (\ref{set2}). The quartic polynomial (\ref{poly}) is then solved for $z$ using the Newton-Raphson method, with a fixed value of $\rho_0$. In this way the equilibrium values of all variables can be found. We have used a total of 40 amplitudes in the Fourier expansion of each $h_i(\phi)$  (we checked that 
this number ensures a strict numerical tolerance in the iterative process). 
Gaussian quadrature is used to calculate the angular integrals. 

From its definition in terms of the total free-energy density $\Phi$, $\displaystyle{\beta p=\eta \frac{\partial 
\Phi}{ \partial \eta}-\Phi}$, the pressure is calculated as 
\begin{eqnarray}
  \beta p=\frac{\rho}{1-\eta}+\frac{\sum_{i,j}\langle\langle A_{ij}^{(\rm spt)}\rangle\rangle}{(1-\eta)^2}.  
\end{eqnarray}
The equation of state (EOS) in reduced dimensionless units, $p^*\equiv \beta pa_0$, will be plotted in Sec. \ref{results} for a particular set of model parameters to compare the results from our model with MC simulations.

\begin{figure}
\includegraphics[width=3.in]{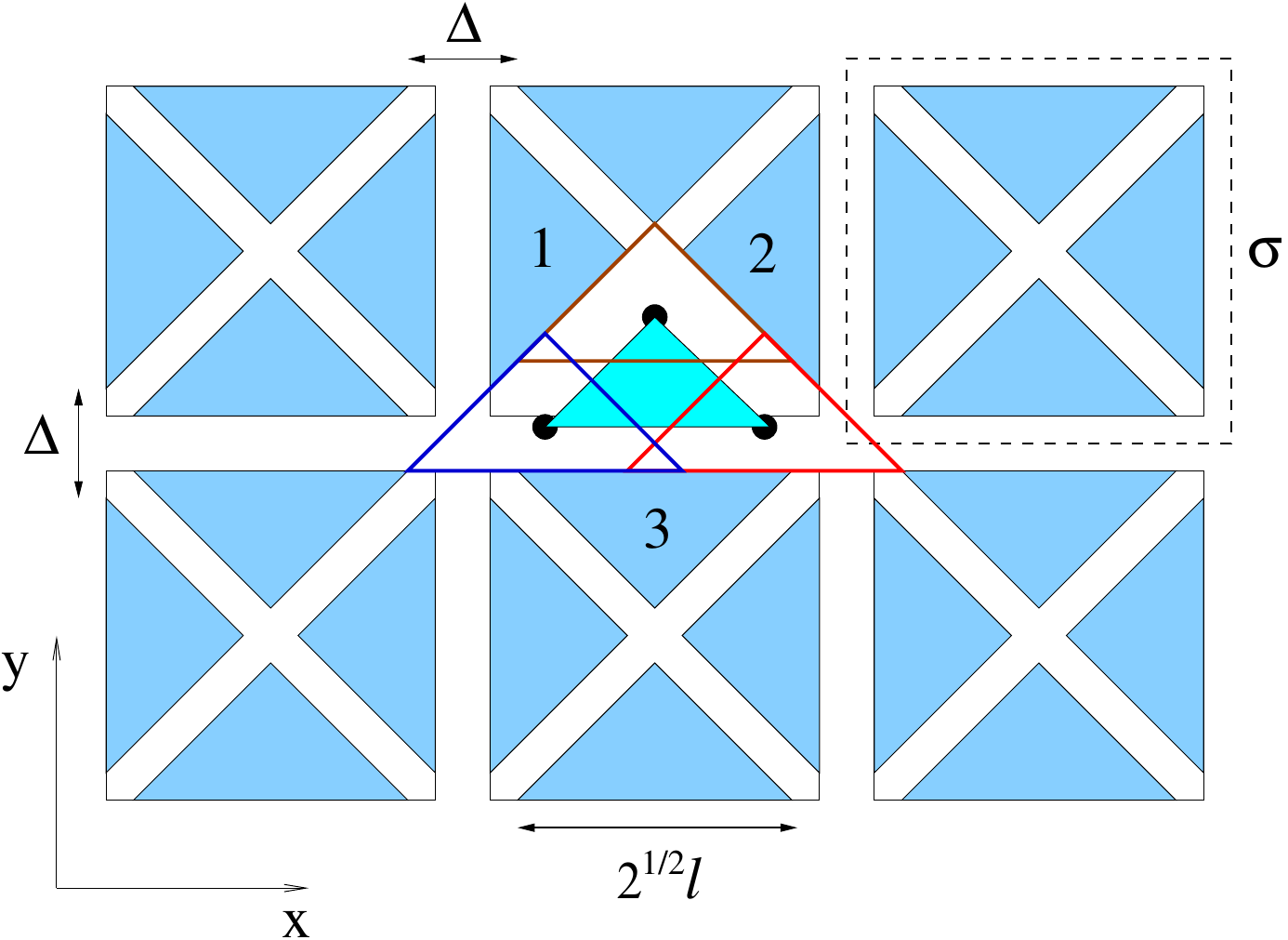}
	\caption{Sketch of right triangles forming a perfect T crystal. The sizes of triangles and the parameter
        $\Delta$ characterizing the crystal are indicated, together with an effective square tetramer of
	size $\sigma=\sqrt{2}l+\Delta$ (the lattice parameter). 
	A particular triangle can be in contact with two of its three neighbours (labeled as 1, 2 and 3) in three different configurations. The perimeter of the free area region, and thus its area, can be obtained by 
	joining the center of mass of the triangle in these configurations.}
	\label{fig4}
\end{figure}

\subsection{Cell theory}

A cell theory can be used to describe the free energy of the
T crystal phase. As shown later, the theory can describe the T liquid-crystal phase almost quantitatively.
Fig. \ref{fig4} shows the particle configurations of a perfectly orientationally ordered
T crystal of right triangles characterized by the parameter $\Delta$ (distance between vertexes of neighbor triangles along $x$ and $y$-directions).
It is easy to check that the free area for the center of mass of one
triangle constrained to move inside the cavity formed by their neighbors also
has a right triangular geometry,
with base and height equal to $2\Delta$ and $\Delta$, respectively. The 
free area is then $A_{\rm free}= \Delta^2$. Taking the
unit cell as a square with lattice parameter 
$\sigma=\sqrt{2}l+\Delta$ ($l$ being the common length of the two 
identical sides),
inside which four triangles are found, the packing fraction of the
crystal is obtained as $\eta=4a_0/\sigma^2$, which gives
\begin{eqnarray}
\eta=\left(1+\frac{\Delta}{\sqrt{2}l}\right)^{-2}.
\end{eqnarray}
The free energy per particle of the CeT approximation, in reduced thermal units,
is $\varphi=-\log\left(A_{\rm free}/\Lambda^2\right)$ (with $\Lambda$ the thermal length). Expressing $A_{\rm free}$ as a function of $\eta$,
\begin{equation}
        \varphi=-2\log\left[\frac{\sqrt{2}{l}}{\Lambda}\left(\eta^{-1/2}-1\right)\right],
	\label{f_CeT}
\end{equation}
The pressure is obtained by differentiating $\varphi$ with respect to $\eta$.
In reduced units,
\begin{equation}
p^*=\eta^2\frac{\partial \varphi}{\partial \eta}=\frac{f_s\eta}{1-\sqrt{\eta}},
	\label{pressure}
\end{equation}
where $f_s=1$.
This expression is identical to that of a fluid parallel hard squares in a 
simple square lattice. This is not simply a coincidence, since 
the 2D-periodic extension (with period $\sigma$ along $x$ and $y$) of square 
tetramers (shown in Fig. \ref{fig4} with solid lines) is equivalent to
a simple square lattice of parallel squares.

The CeT expression (\ref{pressure}) explicitly assumes that the orientation of the particles is frozen.
In fact this is not the case, the degree of orientation being described quantitatively by a set of order parameters.
An extension of the CeT approximation consists of allowing for the central particle to orient along different
directions, still keeping the orientation of its neighbors fixed. The calculation can be done easily using MC integration.
In turn, two situations can be implemented for the orientations: equal probability, or orientations weighted by the
equilibrium distribution obtained from simulation. These two limits provide corrections to 
Eqn. (\ref{pressure})
as weakly density-dependent amplitudes $f_s$ that range from $\sim 1.5$ in the equal-probability case to $\sim 1.2$ when the
equilibrium distribution is used to sample the orientations. In the following we set $f_s=1.5$, in the understanding
that this is a phenomenological correction (we must note that the true structure of the tetratic crystal is chiral and
involves local rotations of the triangles, a feature that is not incorporated in the CeT approximation).


\subsection{Distribution functions and order parameters}

Once the equilibrium angular distribution functions $\{h_i(\phi)\}$ for all clusters are found, the corresponding equilibrium angular functions of monomers (free ones and those that making up different clusters) can be calculated. We enumerate clusters 
as follows: cluster 1 corresponds to free monomers, and their contribution to the total angular function is 
$\displaystyle{h_1^{(m)}(\phi)=\frac{\rho_1}{\rho_0}h_1(\phi)}$. We should remember that $h_1(\phi)$ quantifies the probability density for the triangle axis (pointing from the base to the $90^{\circ}$-vertex along the height) to be oriented with respect to some fixed reference axis at an angle $\phi$. Note that $h_1(\phi)$ contains a factor $\rho_1/\rho_0$, the fraction of monomers. The second cluster 
is the triangular dimer, made from two monomers, each pointing at angles $3\pi/4$ and $-3\pi/4$ with respect to the dimer triangular axis. Then 
$\displaystyle{h_2^{(m)}(\phi)=\frac{\rho_2}{\rho_0}\left(h_2(\phi+3\pi/4)+h_2(\phi-3\pi/4)\right)}$. The third cluster corresponds to one of the rhomboids, made again of two monomers 
oriented at angles $\pi/4$ and $-3\pi/4$ with respect to the rhomboidal axis (parallel to the contact edge). Then $\displaystyle{h_3^{(m)}(\phi)=\frac{\rho_3}{\rho_0}
\left(h_3(\phi+\pi/4)+h_3(\phi-3\pi/4)\right)}$. The other rhomboidal cluster is the mirror reflection of the latter, so we have 
$\displaystyle{h_4^{(m)}(\phi)=\frac{\rho_4}{\rho_0}\left(h_4(\phi-\pi/4)+h_4(\phi+3\pi/4)\right)}$. The square dimer (the 5th species) if made of two 
monomers with axes parallel and antiparallel to any of the diagonal of the square,
so we have $\displaystyle{h_5^{(m)}=\frac{\rho_5}{\rho_0}\left(h_5(\phi)+h_5(\phi+\pi)\right)}$. 
Finally,  we have the square tetramers (the 6th species), made of four monomers at angles 
$\pm \pi/4$ and $\pm 3\pi/4$ with respect to the square diagonals, and consequently 
 $\displaystyle{h_6^{(m)}(\phi)=\frac{\rho_6}{\rho_0}\sum_{k=\pm 1}\left(h_6(\phi+k\pi/4)+h_6(\phi+3k\pi/4)\right)}$. The total angular function of monomers is simply the sum 
 \begin{eqnarray}
 h(\phi)=\sum_{i=1}^6 h_i^{(m)}(\phi) 
 \end{eqnarray}
 From this, the order parameters 
 of all the monomeric units can be calculated:
 \begin{eqnarray}
     Q_{2n}=\int_0^{2\pi} d\phi \cos(2n\phi)h(\phi),\quad n=2,4.
 \end{eqnarray}
This angular function and the order parameters represent the degree of ordering of the HRT fluid and will be shown later. We note that the T directors can change from the original ones when the relative compositions of clusters change due to the shift of the main peaks of $h(\phi)$ by $\pm \pi/4$. This could change the sign of $Q_4$. In the following we will always plot the absolute value $|Q_4|$ when necessary. 

\subsection{Choice of association energy}
\label{parameters}

The values of the parameters that define the association energy (see Eqn. (\ref{association})) were chosen as follows. Note that for free monomers we always set $\epsilon_1^{(1,2)}=0$. In some cases, we also set $\epsilon_i^{(2)}=0$ for any $i$, while 
we considered a repulsive association energy for $\epsilon_i^{(1)}$, i.e. a penalty barrier that prevents the formation of clusters:
\begin{eqnarray}
\epsilon^{(1)}_2=\epsilon_3^{(1)}=\epsilon_4^{(1)}=\epsilon>0,\hspace{0.3cm}
\epsilon_5^{(1)}=\sqrt{2}\epsilon,\quad \epsilon_6^{(1)}=4\epsilon.
\end{eqnarray}
Therefore, for all dimers joined by the catets this penalty is the same, while for the square dimer joined by the triangle hypotenuse the penalty is a factor of $\sqrt{2}$ larger. Finally the penalty for the square tetramer, with four catet contacts, is 4 times larger. In summary, the penalty energy is taken to be proportional to the total contact length of the particles. The coupling with the T order parameter is taken in some cases by selecting
\begin{eqnarray}
\epsilon_i^{(2)}=\delta<0
\label{delta}
\end{eqnarray}
for any $i\neq 1$, which means that the T ordering enhances the formation of all clusters in an equivalent way for any $i\neq 1$. In principle a more meaningful 
choice is to make $\epsilon_i^{(2)}$ to depend on $i$ because square 
dimers and tetramers are certainly more abundant than the rest of clusters 
in the O and T phases. However 
we decided for simplicity to use a single coupling parameter for all clusters.

As shown in Sec. \ref{results}, excluded-area interactions alone predict the system to be mainly composed of tetramers even at moderate packing fractions. This is not seen in simulations, where tetramers appear to grow at high $\eta$. This in turn can be understood in terms of excluded-area interactions: As we consider tetramers to be perfect squares (monomers are in contact along their edges), the total entropic interaction between all clusters promotes a huge clusterization to the geometry with the largest loss in excluded area, which is the square geometry made from four monomers. In simulations clusters can be seen not to be perfect, and they persist during a finite lifetime. Thus, excluded-area interactions alone overestimate the clusterization of the fluid, something that can be prevented by selecting the values of $\epsilon_i^{(1)}>0$ to be positive, i.e. creating a penalty barrier for the clusterization which is higher in the case of tetramers. Also, simulations show that clusterization is strongly promoted when the T orientational ordering is high, and consequently the values of $\epsilon_i^{(2)}<0$ should be negative. In principle the $\epsilon_i^{(j)}$ parameters should depend on packing fraction $\eta$. However, for simplicity, they are considered to be constants.

Finally, to measure the degree of fractionation of monomers among different clusters, we use the area fraction of cluster $i$: 
\begin{eqnarray}
\nu_i=\frac{\rho_i n_i}{\rho_0}.    
\end{eqnarray}

\begin{figure}[h]
	\includegraphics[width=2.5in]{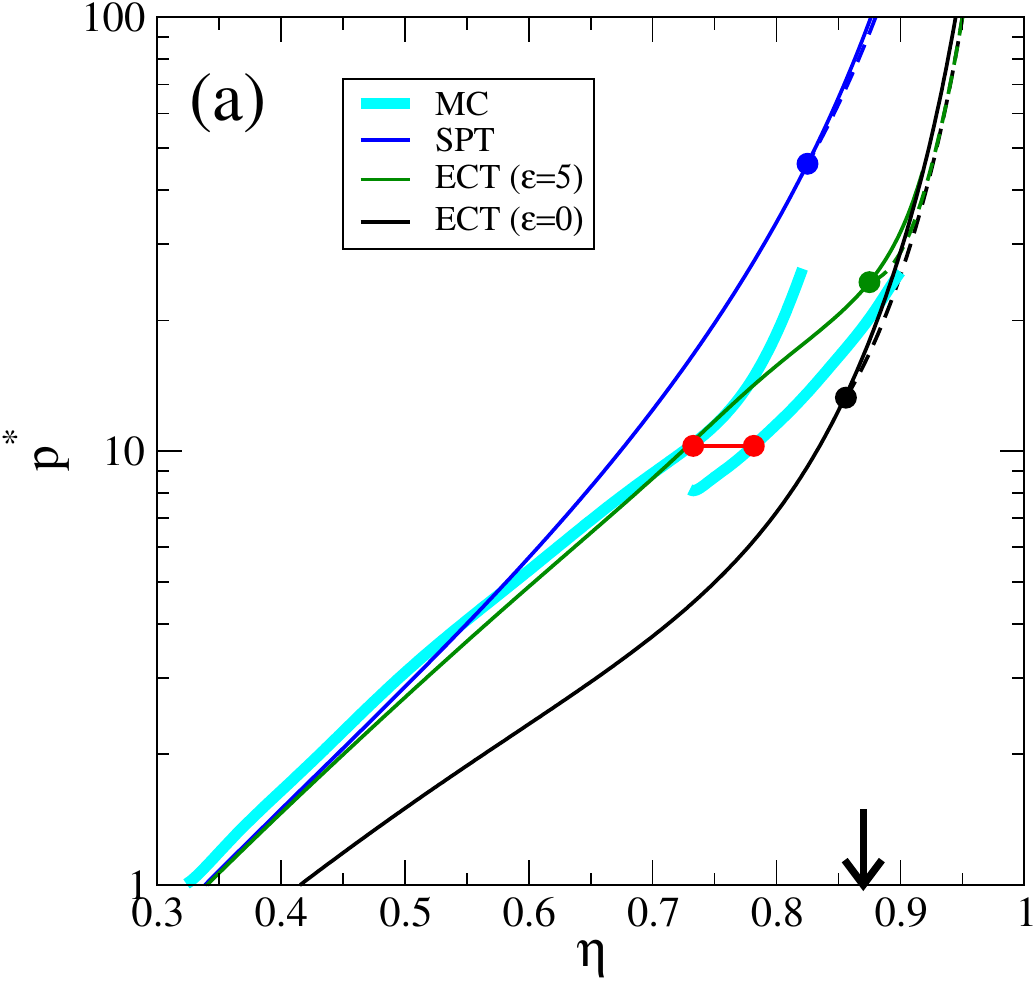}
	\includegraphics[width=2.5in]{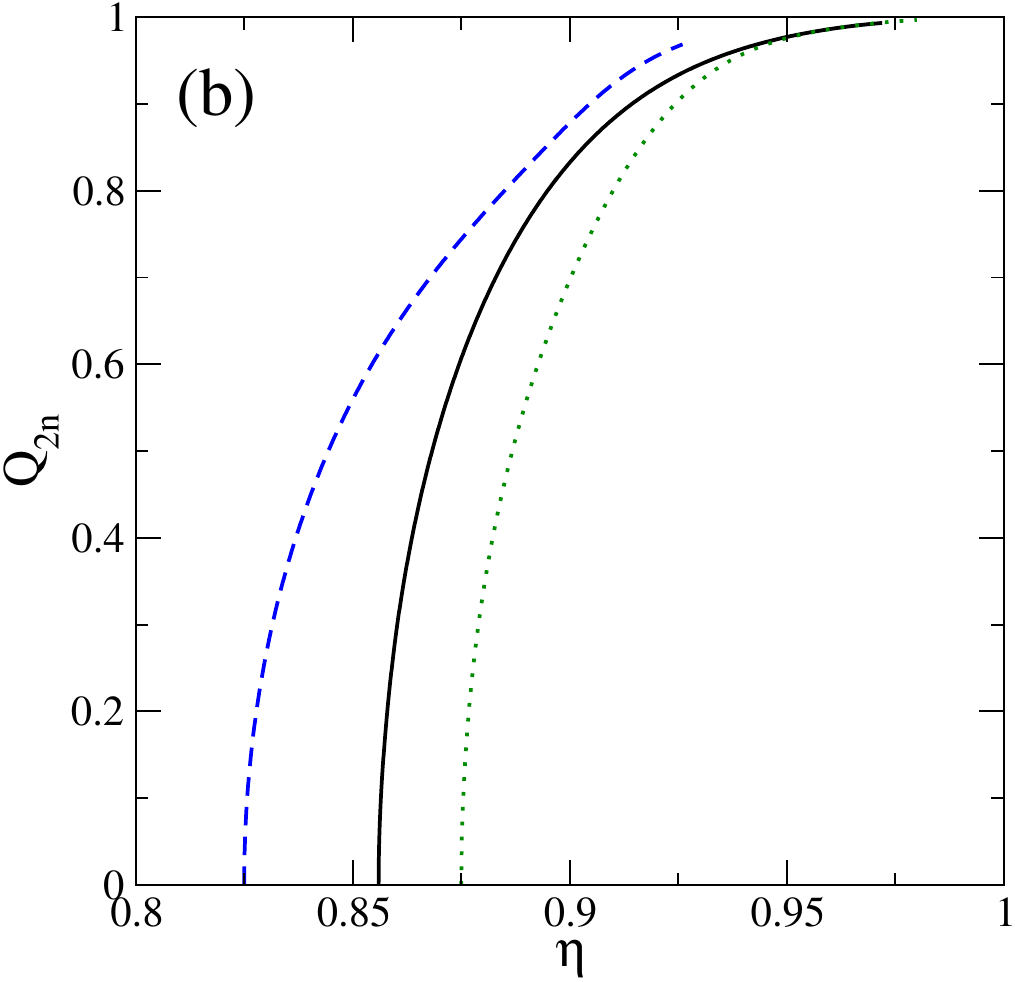}
	\caption{\label{fig5} (a) EOS obtained from SPT, ECT with $\delta=0$ and $\epsilon=0$ or $5$, and MC simulations 
	(data extracted from Ref. \cite{GAN}). The I branches are shown by solid lines, while dashed lines represent the N (from SPT) or T (from ECT) 
	branches. Filled circles indicate the bifurcation points from the I to N (SPT) and T (ECT) phases. Filled circles joined with a line indicate the I-T first order transition 
	from MC. 
	The arrow indicates the location of the T-crystal transition, as obtained from MC simulations \cite{GAN}.
	(b) N (dashed) and T (solid and dotted) order parameters of the monomeric
    units as obtained from the SPT (dashed), ECT with $\epsilon=\delta=0$ (solid) and 
    $\epsilon=5$, $\delta=0$ (dotted).}
\end{figure}

\section{Results}
\label{results}

\begin{figure}[h]
	\includegraphics[width=2.5in]{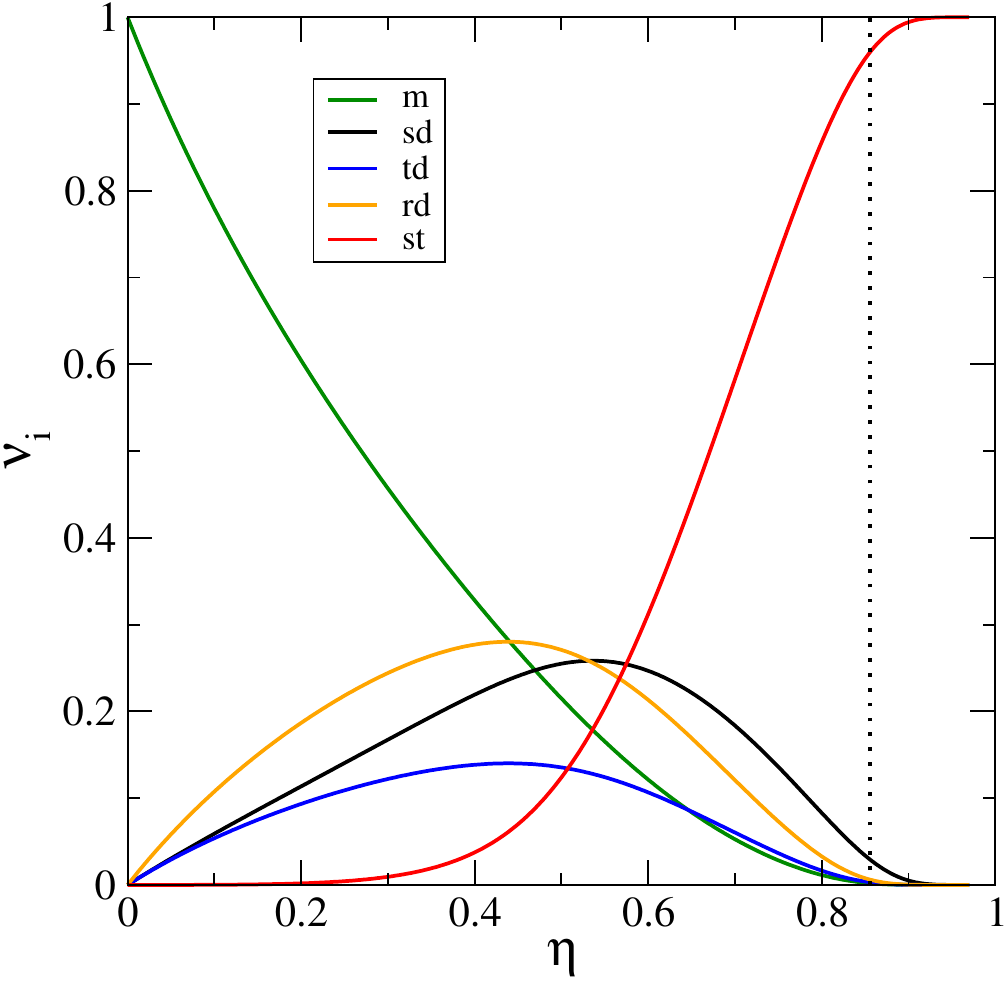}
	\caption{\label{fig6}Area fraction composition of monomers (m), square dimers (sd), triangular dimers (td), rhomboidal dimers (rd) and square tetramers 
	(st) of the I phase (those of the T phase are indistinguishable form the I ones) as a function of the monomer packing fraction 
	obtained from ECT with $\epsilon=\delta=0$. 
	The vertical dotted line shows the I-T second order transition.}
\end{figure}

We first refer to Fig. \ref{fig5}(a), which shows the equation of state $p=p(\eta)$ as obtained from SPT and from the new ECT theory, 
using two values for the association energy. MC results \cite{GAN} are also included for comparison. As can be seen from the figure, 
the SPT greatly overestimates the pressure, predicting a bifurcation from the I to the N phase at a density which is too high. 
Fig. \ref{fig5}(b) shows the behavior of the uniaxial order parameter $Q_2$ (dashed line) along the N branch, starting at the I-N 
bifurcation point. Even worse, the oriented fluid is predicted to be a uniaxial phase, while the MC simulations predict a T (4-atic) phase. 
Note that the small difference between the EOS from SPT (based on the second virial coefficients) and MC simulations at 
$\eta\approx 0.3$ implies that virial coefficients higher than the second are necessary to better describe the EOS even at these low densities.   

The results change significantly by considering a mixture of clusters, still without the association free-energy contribution. In this case the pressure is severely underestimated at low density, but it seems to nicely converge to the pressure of the T (either fluid or solid) phase at high density. The tetratic order parameter $Q_4$ (measuring the ordering of the correct liquid-crystal symmetry) is shown in Fig. \ref{fig5}(b). The predicted bifurcation is still too high as compared with simulation. Finally, when an association term is included, with $\epsilon_i^{(2)}=0$ but $\epsilon_i^{(1)}>0$ (so that association into clusters is penalised), the pressure improves substantially in the whole density interval, and the theory correctly identifies the T phase as the equilibrium oriented fluid phase. However, the location of the bifurcation still occurs at a high density, and in fact it gets worse with respect to the previous theory (see dotted line in Fig. \ref{fig5}(b) which represents the $Q_4$ order parameter).

Fig. \ref{fig6} shows the fraction of clusters as a function of density for the mixture theory. The vertical dotted line indicates the location of the bifurcation to the T phase. The monomer fraction decreases with density, while the fraction of dimers increases. Note that the two enantiomers of the rhombic dimer clusters (Fig. \ref{fig1}) appear with equal probabilities, making the I phase achiral. At high density the fraction of tetramers increases strongly, while the rest of clusters appear in less and less proportions, so that at bifurcation the system almost completely consists of tetramers. However, as shown in Fig. \ref{nuevafig2} \cite{MAR3}, the monomer association into clusters begins to be important at densities much higher than shown in Fig. \ref{fig6}. The increased stabilisation of clusters may explain the dramatic lowering of pressure with respect to simulation. We will see below that the association term corrects this behavior quite substantially.

\begin{figure}[h]
	\includegraphics[width=2.5in]{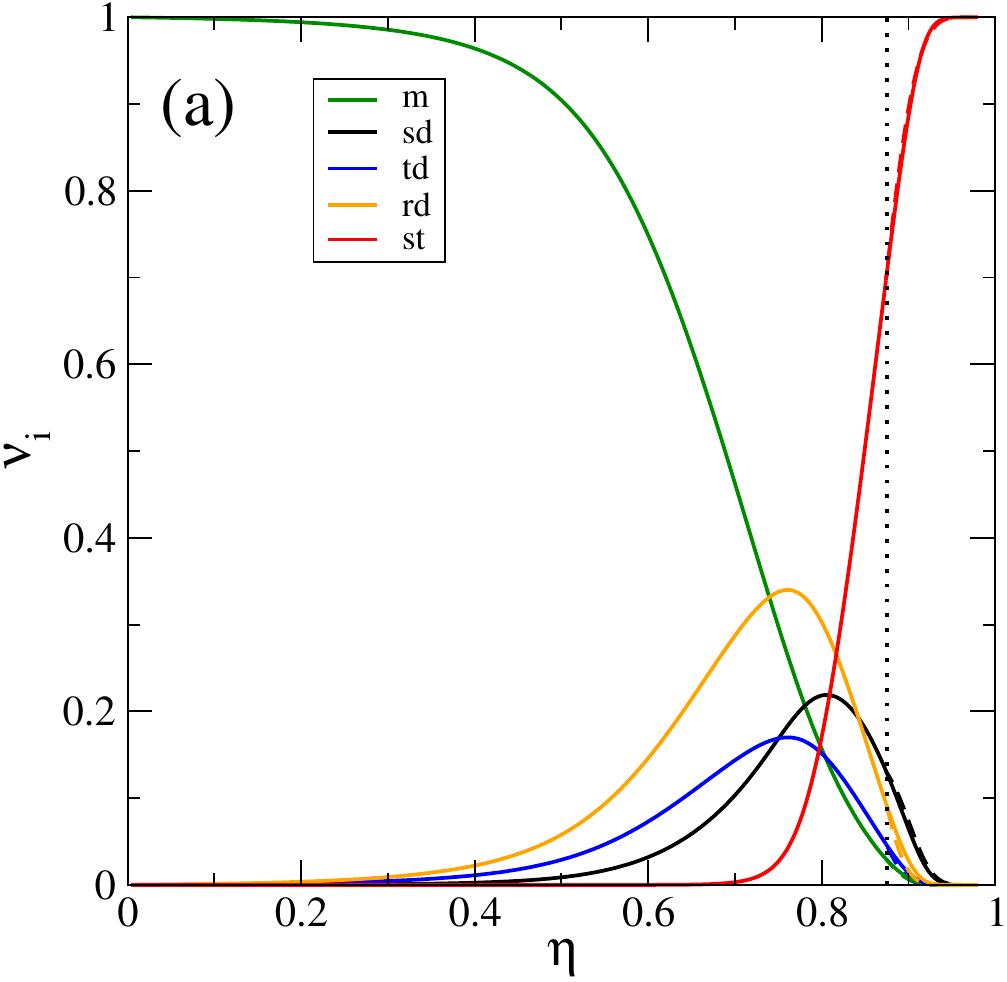}
	\includegraphics[width=2.5in]{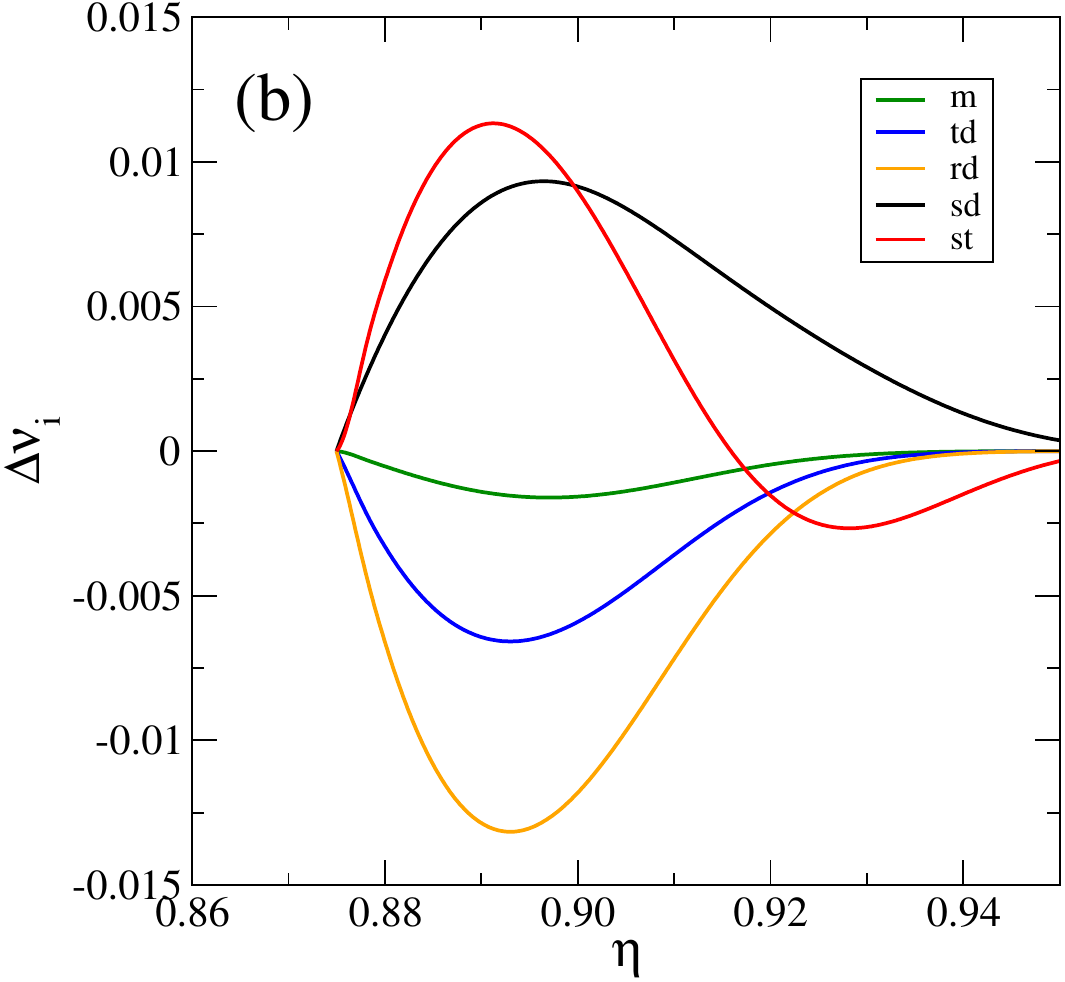}
	\caption{(a) Area fraction compositions of monomers and different clusters (as they are labeled) as a function of the monomer packing fraction obtained 
	from the ECT with $\epsilon=5$, $\delta=0$. With dashed lines (hardly visible) we show the area fraction compositions of the T branch. The dotted vertical line 
	indicates the I-T second order transition. (b) Differences between the area fraction compositions of the T and I phases of monomers and 
	different clusters.}
 \label{fig7}
\end{figure}

Fig. \ref{fig7}(a) shows the area fraction of clusters (solid lines) as a function of packing fraction for the I phase, when the repulsive energy is set to $\epsilon=5$. In this case the fraction of monomers stays at a large value up to $\eta=0.4$, from which it begins to decrease. Fractions of the other clusters then begin to increase and reach a maximum in the density interval $0.7-0.8$, decreasing for larger densities. It is then that the fraction of tetramers grows and finally saturates to unity for very large packing fractions. We note that this behavior is qualitatively similar to that observed in the simulations \cite{MAR3}, see Fig. \ref{nuevafig2}. Also note that the fractions of the two types of rhomboidal clusters are the same, so that the system remains achiral.

In Fig. \ref{fig7}(b) the difference in cluster area fractions between the T and I phases (i.e. T minus I) is plotted at the I-T bifurcation density and beyond. The fraction of free monomers, triangular dimers and rhomboidal dimers is lower in the T phase, as opposed to square dimers and tetramers which are much more abundant (except for $\eta>0.917$). This is understandable since square clusters tend to stabilize the T order (note that at high density the square tetramers are the one ones present in the fluid).

\begin{figure}[h]
	\includegraphics[width=2.5in]{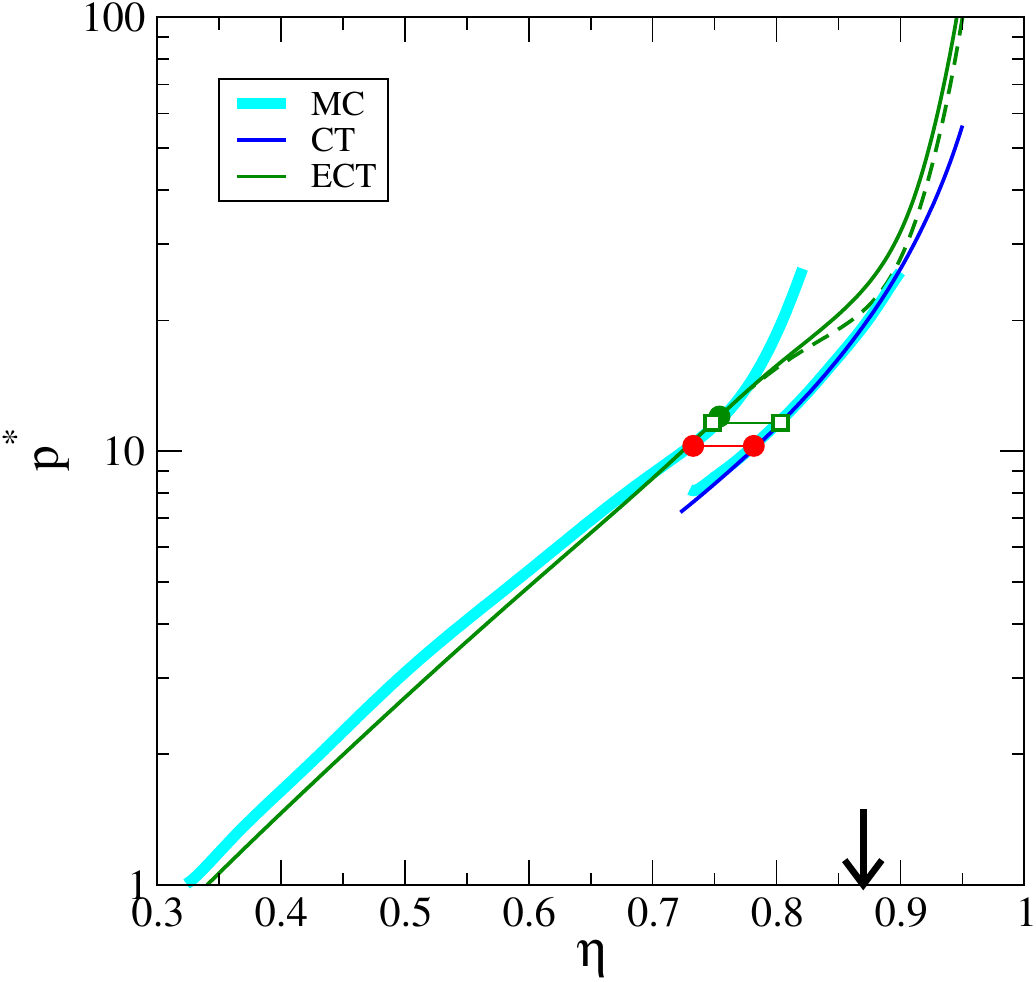}
	\caption{EOS from the ECT (with $\epsilon=5$ and $\delta=-0.88$) for I (solid line) and T (dashed line) phases, and from the CeT for the T phase (taking $f_s=1.5$). 
	The MC results are also shown (data extracted from Ref. \cite{GAN}). The 
	filled circle indicates the I-T bifurcation given by the ECT. Filled circles joined by a line indicate the coexistence packing fraction values of the I-T transition obtained from MC simulations. Open squares joined by a line indicate the coexistence values as obtained from the I branch of the ECT and the T-phase 
	branch obtained from the CeT (with $f_s=1.5$). The arrow indicates the T-crystal transition, as obtained from MC simulations \cite{GAN}.} 
	\label{fig8}
\end{figure}

We now consider the coupling between the tetratic order parameter and the association energy. This coupling promotes  cluster formation, see Eqns. (\ref{association}) and (\ref{delta}). Keeping the penalty barrier as before ($\epsilon>0$), the I-T bifurcation now occurs at lower densities, compared with the case $\delta=0$. Fig. \ref{fig8} shows the resulting EOS from the ECT with $\epsilon=5$ and $\delta=-0.88$. The I-T second-order transition now takes place rather close to the coexisting value of the I phase at the first-order I-T transition, as obtained from MC simulations \cite{GAN}. The EOS of the T crystal, obtained from the CeT (with $f_s=1.5$), is also plotted in the figure. The agreement between this EOS and the T liquid-crystal branch from MC simulations is remarkable. If we take the I branch obtained from the ECT, and the T branch from the CeT of the crystal phase, and calculate the I-T coexistence, the resulting coexistence density values are very close to those from MC simulation. These results highlight the important role of clustering and spatial correlations in the quantitative prediction of the EOS of the T phase of HRT fluids. 
As seen in Fig. \ref{fig8}, despite the fact that the packing fraction of the I-T bifurcation point is
improved and is now close to the I-T coexisting 
values found in simulations, the EOS for the T phase predicted by the theory in the neighborhood of the transition    
is quite far from the simulation results. The reason for this is the strong spatial correlations of the liquid-crystalline
T phase, which our density functional, based on a one-particle uniform density, cannot adequately describe. By contrast, a simple cell theory assuming perfect spatial order gives a quantitatively correct EOS.

\begin{figure}[h]
	\includegraphics[width=2.5in]{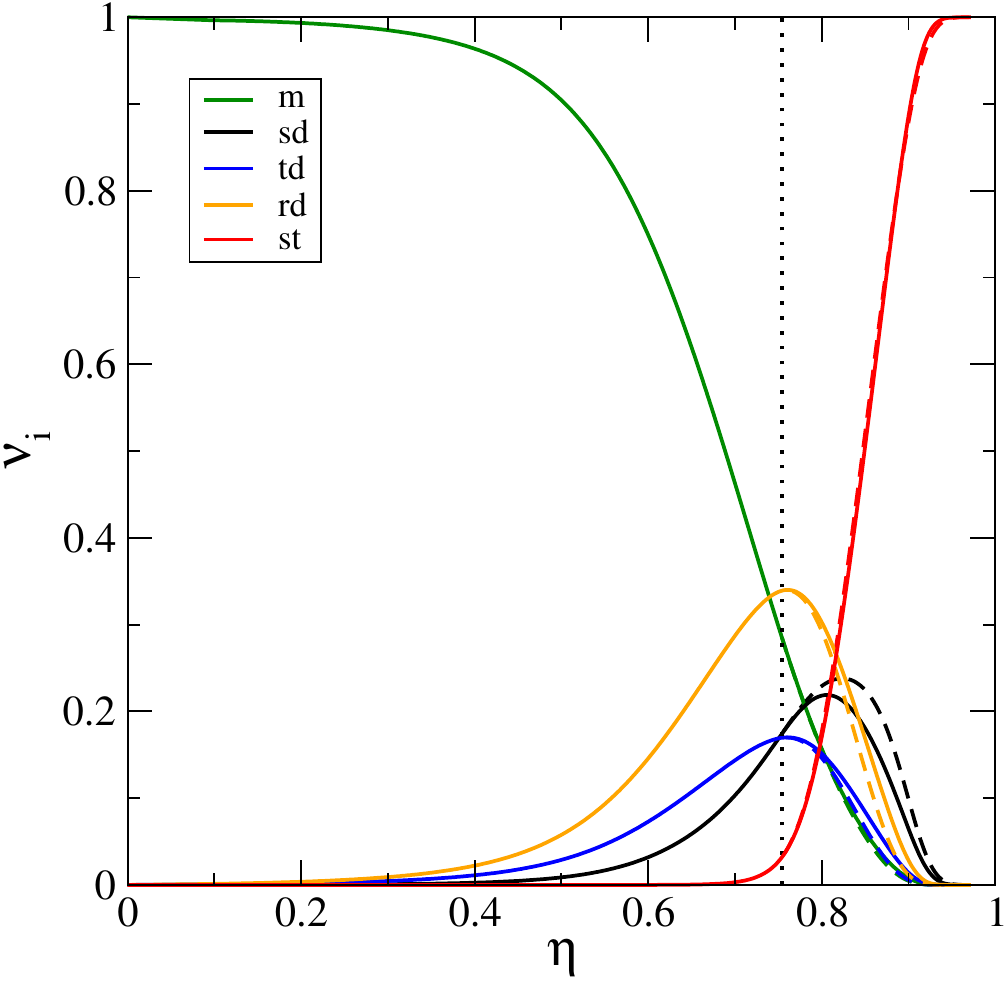}
	\caption{Area fraction compositions from the ECT (with $\epsilon=5$ and $\delta=-0.88$). I and T branches are shown by solid and dashed lines, respectively. The vertical dotted line indicates the I-T second-order transition.}
 \label{fig9}
\end{figure}

\begin{figure}[h]
	\includegraphics[width=2.5in]{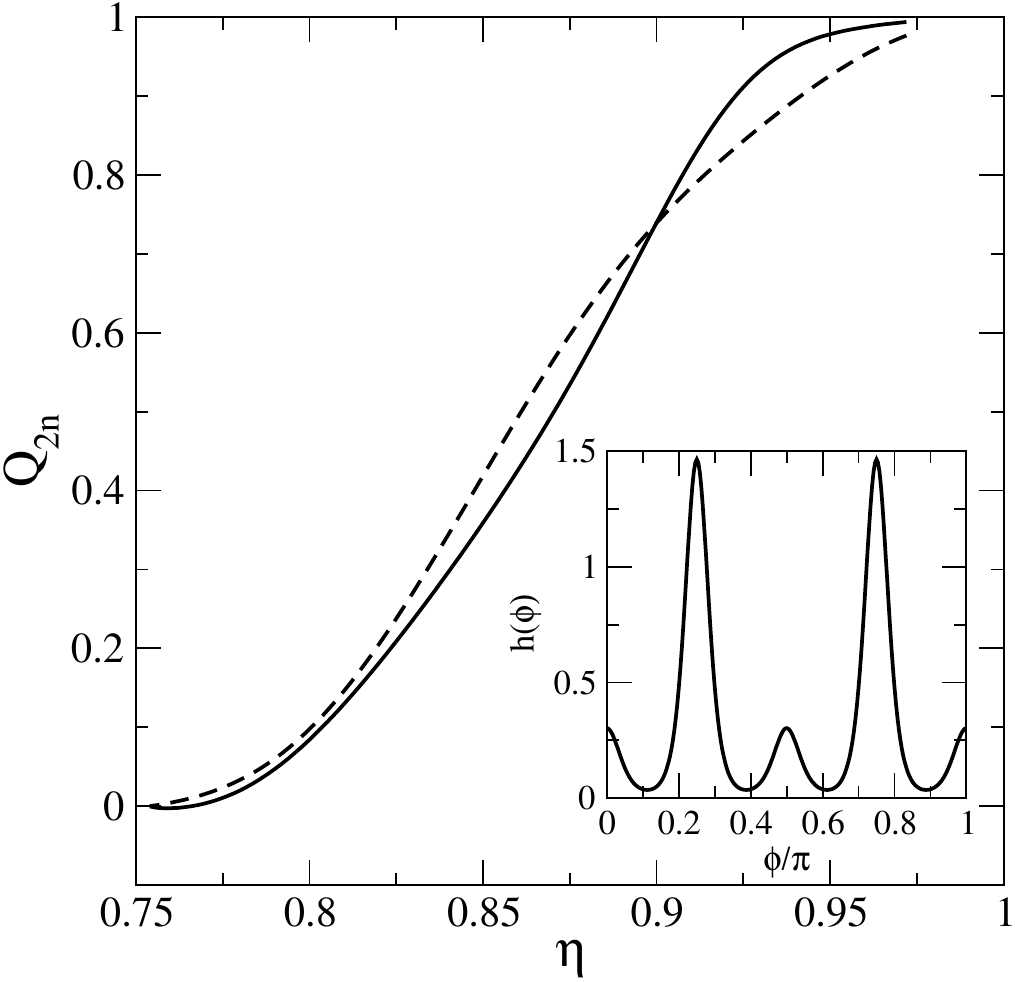}
	\caption{Tetratic ($Q_4$, solid line) and octatic ($Q_8$, dashed line) order parameters of monomeric units plotted from the I-T bifurcation point, as obtained from the ECT with $\epsilon=5$ and $\delta=-0.88$. The inset shows the monomer distribution function for $\eta=0.876$.}
 \label{fig10}
\end{figure}

\begin{figure}[h]
	\includegraphics[width=2.5in]{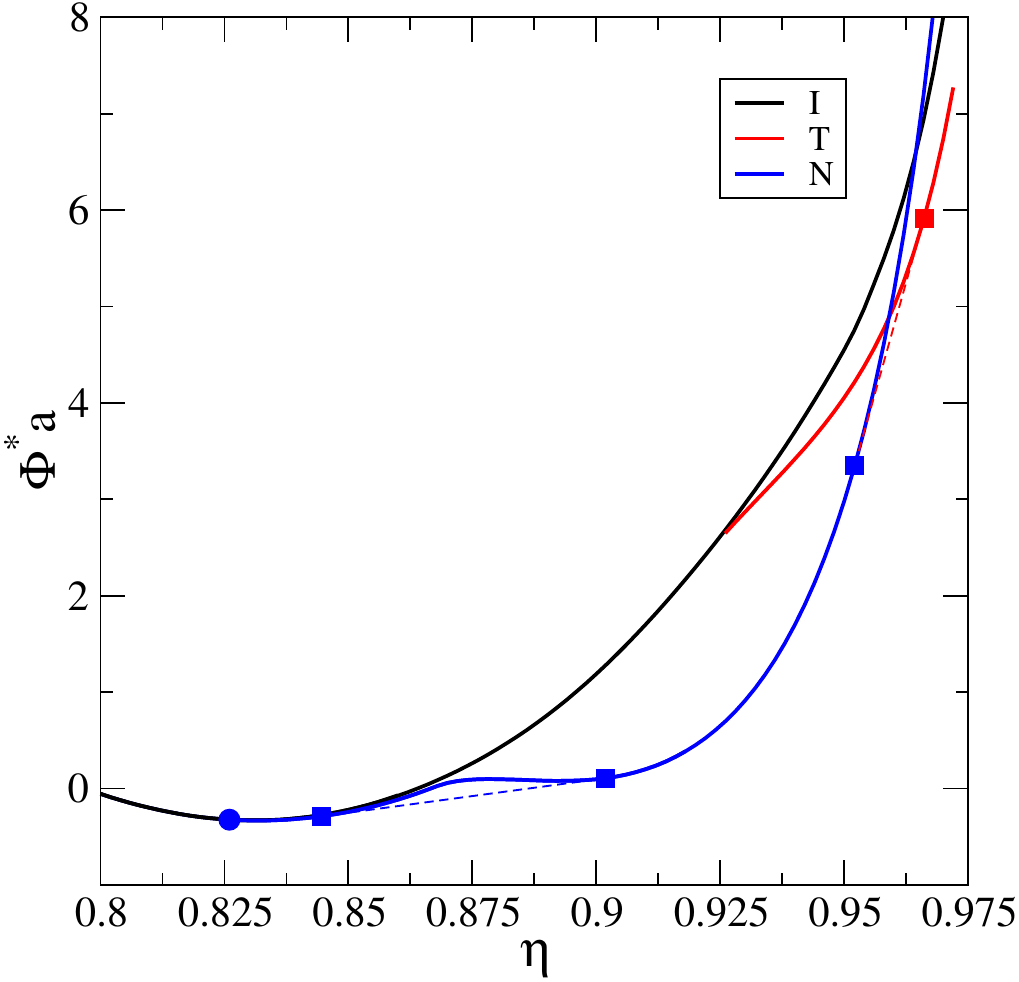}
	\caption{Scaled free-energy density minus a straight line, $\Phi^*a=\Phi a+46.989-65.462\eta$, as a function of the 
	monomer packing fraction for the I, N and T branches, obtained from the ECT with $\epsilon=15$ and $\delta=0$. The filled 
	circle indicates the I-N bifurcation point, while the N-N and N-T two-phase coexistences are indicated by the filled squares. 
	}
 \label{fig11}
\end{figure}

Fig. \ref{fig9} shows the area fraction of clusters as a function of the packing fraction for the same values as above, $\epsilon=5$, and $\delta=-0.88$. The overall trend in similar to that obtained for the values $\epsilon=5$ and $\delta=0$ (see Fig. \ref{fig7}), but with an important difference: slightly above the I-T bifurcation, the fluid is now enriched in all dimers with approximately equal 
proportions, while it is depleted in tetramers. The fraction of the latter abruptly increases with $\eta$ from the bifurcation. From Fig. \ref{fig9} it can be seen that the
T phase has a larger proportion of square dimers compared to the I (metastable) phase (dashed vs. solid lines respectively). Rhomboidal and triangular dimers follow 
the opposite trend (with the I phase being more enriched in them). This behavior 
of the area fraction of monomers as a function of $\eta$ in turn explains why the T order parameter of monomers $Q_4$ has the peculiar behavior shown in Fig. \ref{fig10}. Note that the O order parameter $Q_8$ is higher than $Q_4$ for some values of $\eta$, which is a direct consequence of the presence of satellite peaks located at angles $\pm \pi/4$ off the main peaks of the orientational distribution function of monomers $h(\phi)$ (see inset). This feature is an indication of the importance of O correlations in the system. 

\begin{figure}[h]
	\includegraphics[width=2.5in]{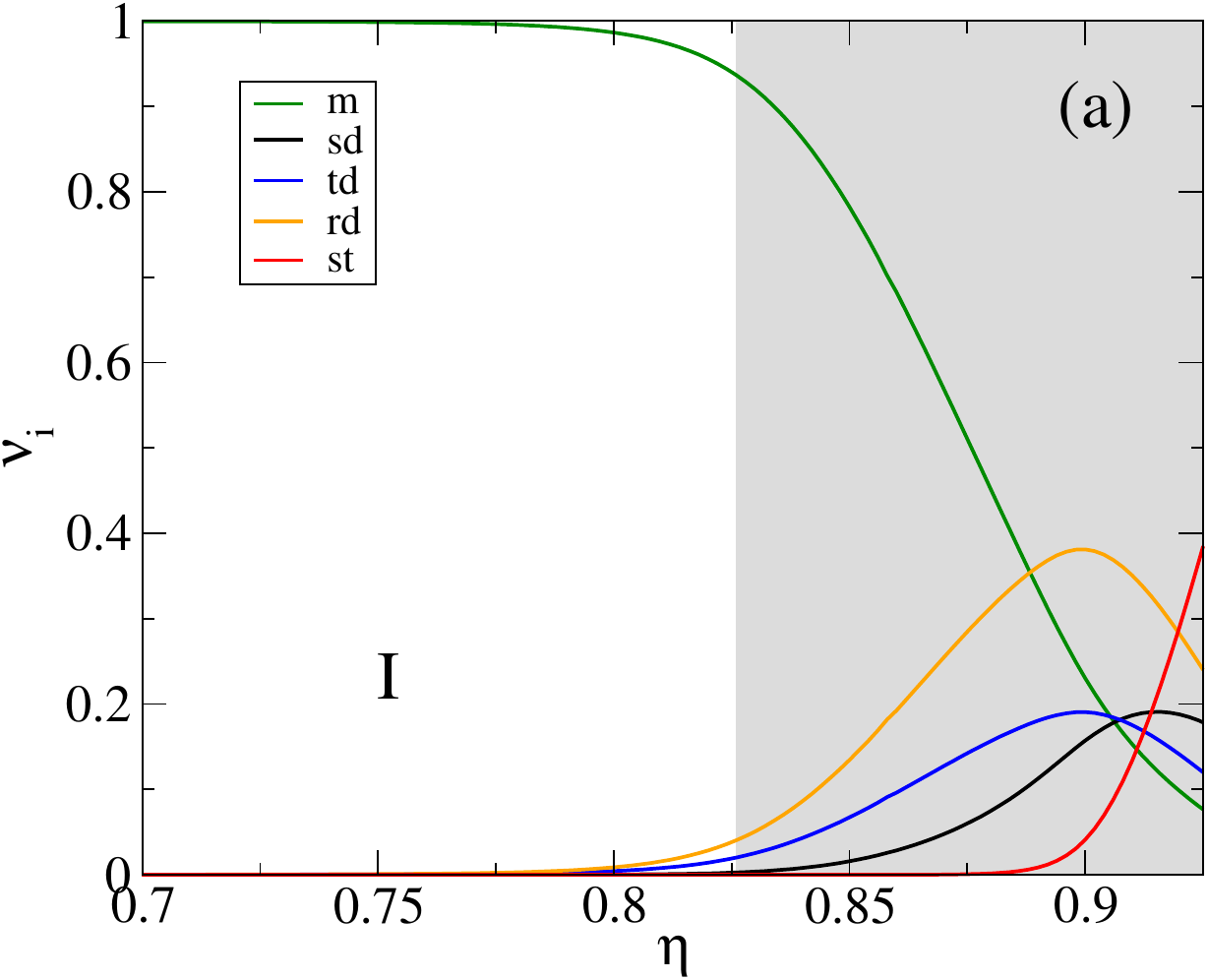}
	\includegraphics[width=2.5in]{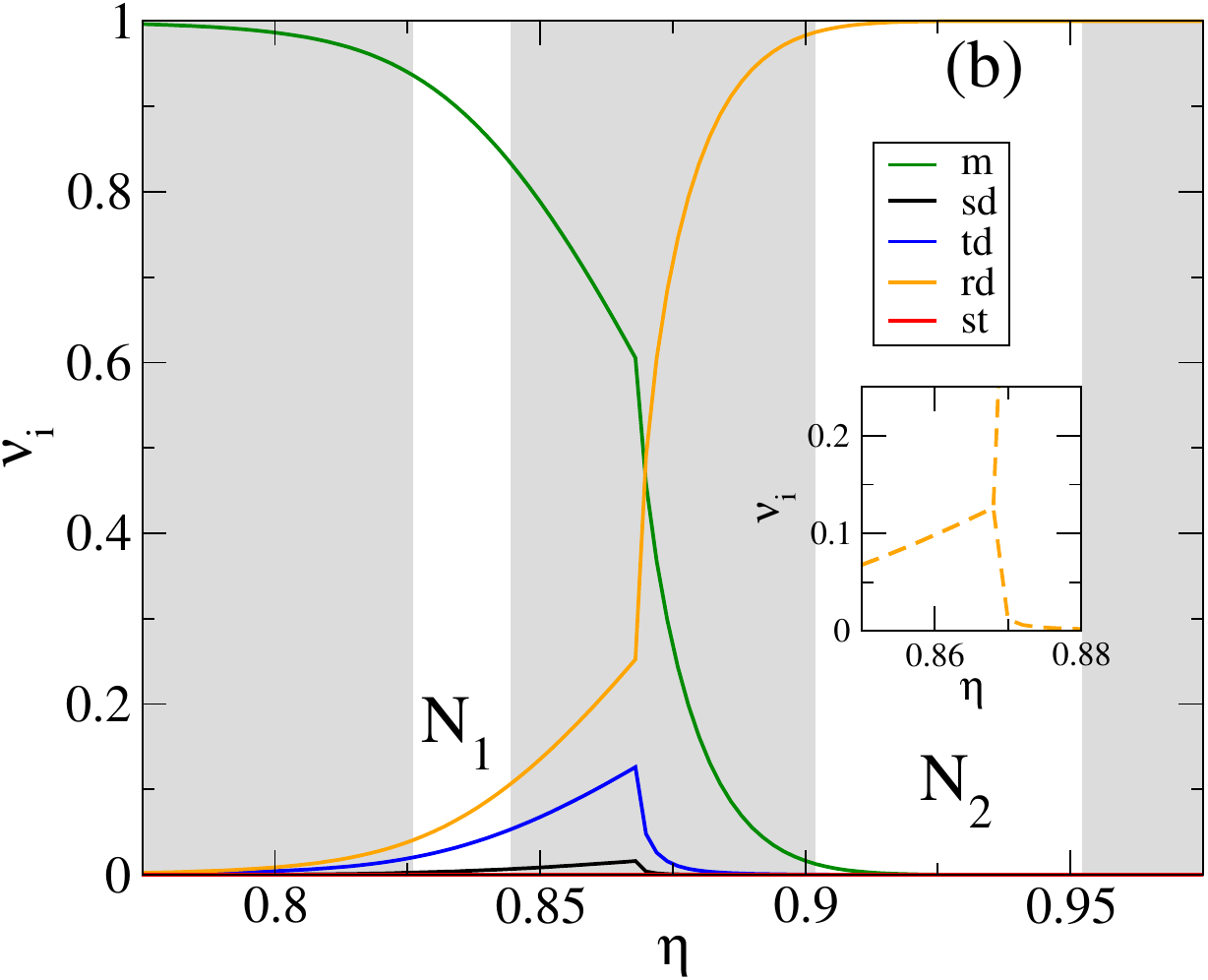}
	\includegraphics[width=2.5in]{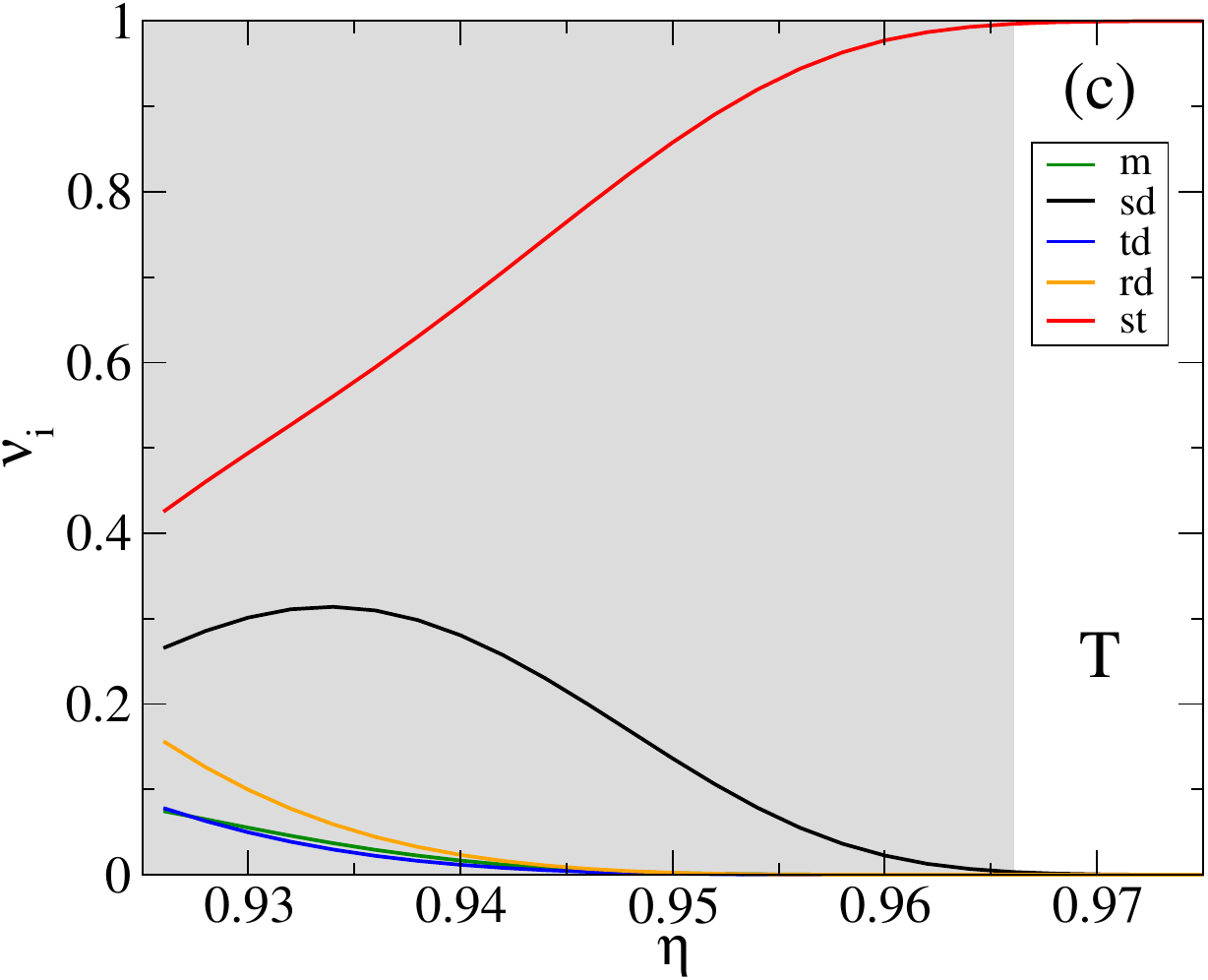}
	\caption{Area fraction compositions vs. packing fraction along the I (a) , N (b) and T (c) branches from the ECT 
	with $\epsilon=15$ and $\delta=0$. The shaded 
	regions indicate the regions of instability of the I (a), N (b) and T (c) phases due to the second or first-order transitions 
	between different phases. The regions of stability of all phases are correspondingly labeled. In panel (b) we 
	show in the inset the area fraction compositions of the two different rhomboidal dimers: at the left of the breaking-symmetry 
	point they are identical while at the right they separate, with one of them eventually dominating as $\eta$ increases.}
\label{fig12}
\end{figure}

At this point the success of ECT to describe the phase behavior of HRT is apparent: (i) It predicts a transition to the T liquid-crystal phase from the I phase at a packing fraction similar to that found in MC simulations. (ii) At sufficiently high densities, the T phase is mainly populated by tetramers, and the ECT can be combined with the CeT to obtain the coexistence densities of the I and T phases, respectively. (iii) Finally, the strong O orientational correlations between particles found in simulations are also predicted from our theory.

However, it is interesting to explore the model in regions of parameter space other than the one which provides the best agreement with MC simulations. In particular we are interested into the phase behavior for large values of the penalty parameter $\epsilon$ and for $\delta=0$, i.e. when the formation of tetramers is greatly suppressed at moderate densities. As discussed below, chiral phases can be found in this regime.

\begin{figure}[h]
	\includegraphics[width=2.5in]{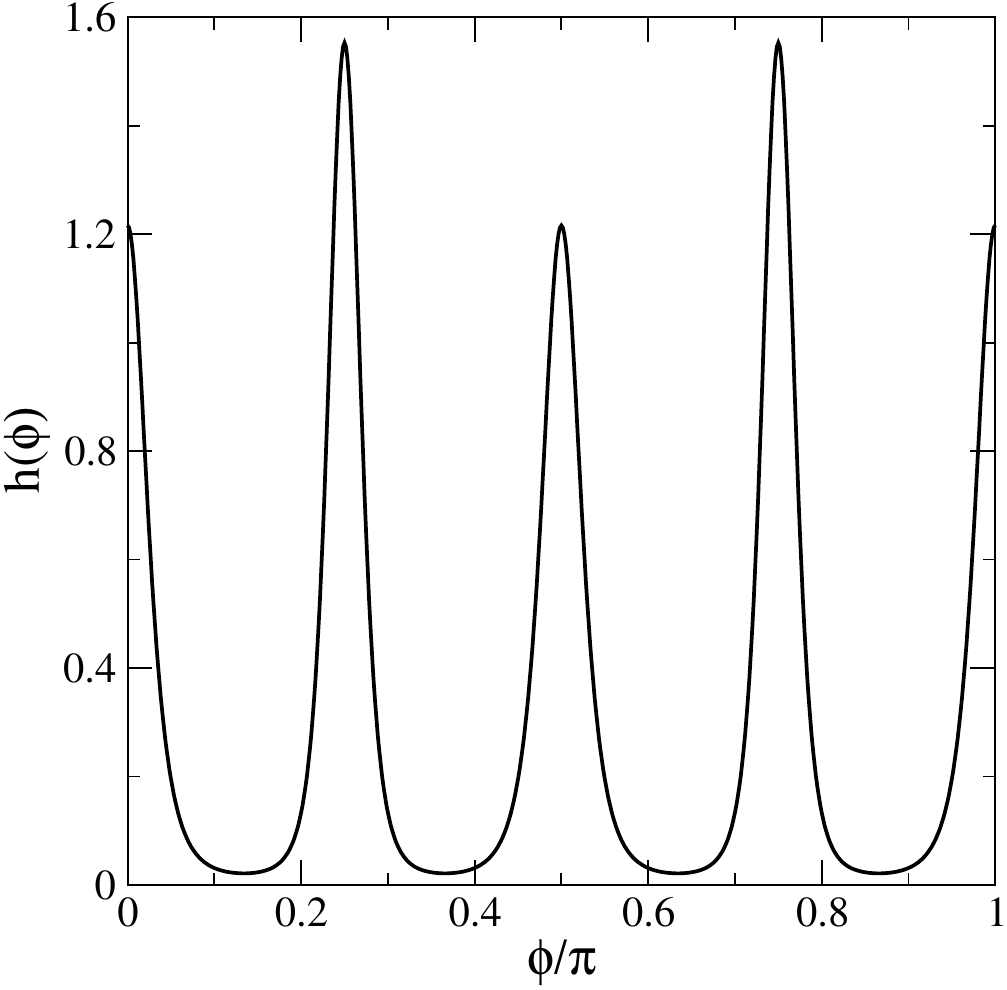}
\caption{
	 Monomer distribution function with quasi-O symmetry obtained from the ECT with the values 
	of $\epsilon=15$ and $\delta=0$ and fixing the monomer packing fraction to $\eta=0.937$.  
	Note that this quasi-O structure is metastable with respect to the 
	N phase.}
\label{fig13}
\end{figure}

\begin{figure}[h]
	\includegraphics[width=2.5in]{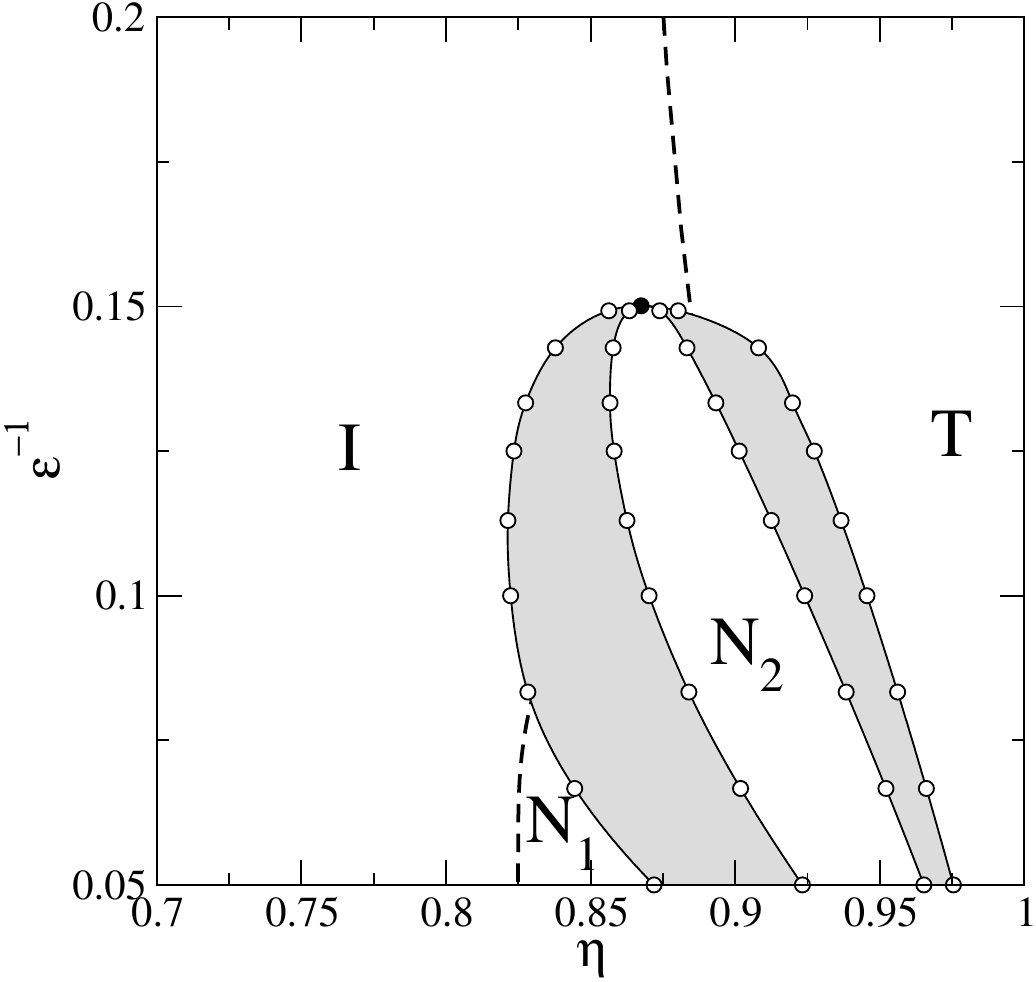}
	\caption{Phase diagram in the $\epsilon$-$\eta$ plane from the ECT with $\delta=0$. The regions of stability of the I, 
	N$_i$ and T are correspondingly labeled. With open circles we show the calculated two-phase coexistences for a 
	certain set of values of $\epsilon$. The 
	filled circle label the azeotropic point. With dashed lines we show the I-N$_1$ and I-T second order 
	transitions. The two-phase coexistence regions are shaded in grey.}
 \label{fig14}
\end{figure}

As an example, Fig. \ref{fig11} shows the free-energy density in reduced units as a function of packing fraction for $\epsilon=15$. Different energy branches are plotted, corresponding to the stable or metastable liquid-crystal phases found. The I phase is stable up to the I-N$_1$ bifurcation point. The $N_1$ phase, mainly populated by triangular monomers, is achiral (the proportion of the two different rhomboidal enantiomers are exactly the same). The stability of this uniaxial phase, predicted by the standard SPT for HRT, has already been discussed in \cite{MAR2,VEL1,MAR}. In this case the phase is stable due to the extremely high proportion of free monomers up to densities $\eta\sim 0.825$ (see Fig. \ref{fig12}). 

Interestingly, further increasing $\eta$ along the N branch, a symmetry breaking point is reached at which the fraction of one of the species of enantiomers increases at the expense of the other species, giving rise to a chiral N$_2$ phase (see Fig. \ref{fig12}). Following the same branch, the N$_2$ phase becomes populated by just one of the enantiomers at high densities, with marginal fractions for the rest of clusters. Fig. \ref{fig11} shows the clear first-order nature of the N$_1$-N$_2$ transition. 

Finally, at very large packing fractions, the N$_2$ phase becomes unstable with respect to the T phase of tetramers, with the N$_2$-T transition being of first order (see Figs. \ref{fig11} and \ref{fig12}). An example of a monomer distribution function with strong O correlations in the T phase is shown in Fig. \ref{fig13}, pertaining to the fluid with $\eta=0.937$. Exactly at this point the T phase becomes unstable with respect to the N$_2$ phase (this example is shown because, as shown in Fig. \ref{fig12}(c), one can design a procedure to obtain a function $h(\phi)$ with this symmetry by combining square dimers and tetramers in similar proportions, with the remaining clusters being marginal in the mixture). 

Fig. \ref{fig14} shows the phase diagram of the model in the $\epsilon^{-1}-\eta$ plane. Phase boundaries were obtained by calculating the free energies of all phases involved. Second-order transitions are indicated by dashed lines, while two-phase coexistence regions are represented by shaded regions. The phase diagram shows that the first order I-N$_1$ and N$_2$-T transitions coalesce into an azeotropic point, above which only a second-order I-T transition remains (this is the phase behavior described before, which corresponds to the values of $\epsilon$ for which good agreement with MC simulations is found).

\section{Discussion and conclusions}
\label{conclusions}

In this work we have proposed a new theoretical framework (ECT), based on density-functional theory, to describe the phase behavior of 2D hard particles, for which standard versions of density-functional theory (such as SPT) fail to predict even the symmetries of the stable liquid-crystal phases. 

The main ingredients of the theory are: (1) Mapping of the one-component fluid onto a multicomponent mixture of different species, consisting of monomers (the original particles) and different  clusters, each formed by aggregations of monomeric units. Clusters are selected to be convex, and one of them should conform with the symmetry of the crystal phase (for example tetramers in the case of HRT). Clusters are obtained by perfectly joining the edges of original particles in different conformations. (ii) Clusters interact through hard-core interactions, so that all the excluded areas between different species are required. These interactions are approximated through the excess part of the free-energy given from SPT. Of course clusters observed in MC simulations are not perfectly shaped, since they form and break with a certain persistence time, which should depend on the packing fraction. As a consequence, our model overestimates the formation of those clusters which tend to minimize the total excluded areas (tetramers in case of HRT). Thus a new term, the association energy, is added to the free energy. This term renormalizes particle interactions to better approximate the real situation. In practice the association energy penalizes the formation of clusters. Also, MC simulations indicate that a coupling between orientational ordering and association energy should exist: for example, T ordering favors the formation of square dimers and tetramers. 
(iii) An important ingredient of the theory is the use of the conservation of the total number of monomers as a constraint. This is taken through a Lagrange multiplier in the free energy. As usual the ideal free energy is taken as that of the multicomponent mixture of species. Minimizing the total free energy, the equilibrium cluster densities and their orientational distribution functions, $h_i(\phi)$, are obtained, and from them all the thermodynamic quantities of the mixture can be derived. (iv) Finally, from the functions $h_i(\phi)$ the corresponding function for monomers (the free ones and those forming clusters), $h(\phi)$, can be obtained by simple geometric relations (these depend on the relative monomer orientations with respect to the main axis of the cluster). In this way the orientational ordering of the original fluid can be quantified. An example was given for the case of HRT. 

Following this procedure, we applied our model to the case of HRT fluid, analyzed the phase behavior, and compared the results with MC simulations. The overall agreement is rewarding: (i) The EOS obtained from the ECT greatly improves upon the standard SPT treatment. (ii) The symmetry of the stable liquid-crystal phase (the T one) is now correctly predicted. (iii) The packing fractions of the I-T phase transition decrease and can be made to be close to those of simulations. (iv) Using CeT to describe the T branch and ECT for the I phase we obtained a first-order I-T transition with values for the coexistence packing fractions in close agreement to those predicted by simulations.

Recently Wittmann et al. \cite{Wittmann} 
have developed a formalism to map soft-matter building blocks that can associate into complexes onto effective
mixtures of particles. The internal degrees of freedom of the complexes can be added as a term in their effective chemical potential. Even though
the theory is developed to describe encapsulation of particles by carriers, it is totally general and can describe aggregation and
clustering. Apart from the trivial use of different ensembles, our theory can be viewed as a particular application of
Wittmann et al. theory to an oriented fluid in which the effective chemical potential is given by a particular dependence on the shape of the
complex (cluster) and the order parameter of the fluid. In fact our choice for the association energy, which is based on  
MC simulation results, rests on a clear physical intuition and can be justified within this formalism.

However, a first-principles development for the association energy is needed to justify our procedure. 
Progress along this avenue is begin made. 
We expect that the present formalism can be applied to any 2D fluid made of polygonal particles, a program that we are actually implementing 
to confirm the success of the ECT in the description of the oriented phases of these fluids. Extension of the theory to 3D 
offers the potential to introduce novel methods for analyzing liquid-crystalline and crystal phases of colloidal systems.

\acknowledgements

Financial support from Grant No. PID2021-126307NB-C21 (MCIU/AEI/FEDER,UE) is acknowledged.


\begin{references}
\bibitem{explicacion} According to the free-energy Monte Carlo calculations of achiral crystal phases presented 
	in Ref. \cite{GAN}, the lowest free-energy crystal consists of a square lattice of square tetramers. It must be said, however,
	that a square lattice of square dimers with random orientations was not analyzed in this study. There are reasons to believe
	that this phase could compete with the phase based on tetramers because of its high configurational entropy.
\end{references}
\end{document}